\documentclass[
    aps,
    prd,
    reprint,
    superscriptaddress,
    amsmath,
    amssymb,
    nofootinbib,
    floatfix
]{revtex4-2}

\usepackage{graphicx}
\usepackage{dcolumn}
\usepackage{bm}
\usepackage{mathtools}
\usepackage{xcolor}

\usepackage[
    colorlinks=true,
    citecolor=blue!70!black,
    linkcolor=blue!70!black,
    urlcolor=blue!70!black
]{hyperref}

\graphicspath{
    {./}
    {figures/paper/}
}

\newcommand{\dd}{\mathrm{d}}

\newcommand{\Order}{\mathcal{O}}
\newcommand{\Lcal}{\mathcal{L}}

\begin{document}


\title{
Uniform Strong-Deflection Lensing in Near-Extremal Kerr
}

\author{Shakibul Chowdhury}
\email{shakibul.chowdhury@gmail.com}

\affiliation{
Department of Physics,
The City College of New York,
City University of New York,
New York, New York 10031, USA
}

\date{\today}


\begin{abstract}

Strong-deflection lensing by a subextremal Kerr black hole exhibits a
logarithmic critical divergence, whereas the exactly extremal prograde
problem develops stronger power-law behavior. We construct a uniform
asymptotic description of the transition between these regimes for
prograde equatorial null geodesics. The distinguished joint
near-critical and near-extremal limit occurs when the impact-parameter
offset from the exact critical value is of order \(M\kappa\), where
\(\kappa=\sqrt{1-a^2/M^2}\). In this limit, the Kerr radial potential
reduces to a quadratic normal form, yielding an analytic crossover
function that connects the subextremal logarithmic and extremal
power-law regimes. We derive a matched next-to-leading-order correction
and validate it against high-precision integrations of the exact Kerr
geodesic equations, finding improvement for all 55 trajectories tested.
The same asymptotic structure yields a higher-order image-branch law
interpolating between algebraic and exponential accumulation, with
crossover order scaling as \(\kappa^{-1}\). We also derive and
numerically validate the corresponding image-branch coordinate-time
delay law, whose high-order extremal-like limit approaches the
horizon-period spacing \(4\pi M\).

\end{abstract}

\maketitle


\section{Introduction}
\label{sec:introduction}

Strong gravitational lensing emerges from null geodesics that approach the critical photon structure of a black hole, undergo large deflections, and subsequently escape toward a distant observer. In this regime, the trajectories acquire a large azimuthal phase and can generate relativistic or higher-order images. The strong-deflection limit was developed systematically for black-hole lensing \cite{VirbhadraEllis2000,BozzaEtAl2001,Bozza2002,Bozza2010}, with subsequent work clarifying its general structure for static, spherically symmetric spacetimes \cite{IyerPetters2007,Tsukamoto2017} and extending the analysis to rotating geometries \cite{Bozza2003,BozzaEtAl2005,BozzaEtAl2006}. For an ordinary unstable photon orbit, the bending angle diverges logarithmically as the impact parameter approaches its critical value, while successive higher-order images accumulate exponentially toward the critical curve \cite{Bozza2002,BozzaEtAl2005,BozzaEtAl2006,IyerPetters2007}.

In Kerr spacetime, the critical null-geodesic structure is richer because frame dragging distinguishes prograde and retrograde photon motion \cite{BardeenPressTeukolsky1972}. The separability of the Kerr geodesic equations and the associated photon-orbit structure have been studied extensively \cite{Carter1968,Wilkins1972,Teo2003, GrallaLupsasca2020Null,CompereLiuLong2022}. The corresponding strong-field optical structure has also been investigated through Kerr ray tracing and caustic analyses, including the behavior of higher-order caustics at high spin \cite{RauchBlandford1994,Bozza2008}. These critical geodesics organize strong-field lensing, the black-hole critical curve, and the hierarchy of increasingly fine image structure \cite{GrallaHolzWald2019,JohnsonEtAl2020, GrallaLupsasca2020Lensing}.

Throughout this work we use geometrized units \(G=c=1\). The
black-hole mass is denoted by \(M\), and the Kerr spin parameter is
\(a=J/M\), where \(J\) is the black-hole angular momentum. We restrict
to the prograde branch \(a>0\), with dimensionless spin \(a/M\).

The prograde near-extremal problem contains an additional singular
limit. As \(a/M\) approaches unity, the unstable prograde critical
orbit moves into the near-horizon region and approaches the event
horizon itself. The resulting high-spin scaling regime has been
studied in connection with extreme-Kerr optical phenomena,
near-horizon photon emission, and critical geodesic motion
\cite{CunninghamBardeen1972,
PorfyriadisShiStrominger2017,
GrallaLupsascaStrominger2018,
LupsascaPorfyriadisShi2018,
KapecLupsasca2020,
GatesHadarLupsasca2021,
YanEtAl2021,
OgasawaraIgata2022}.

At exact extremality, the prograde strong-deflection problem differs
qualitatively from the standard fixed-subextremal case. The special
behavior of photon motion in the extremal and near-extremal Kerr
geometry has been explored in several complementary settings
\cite{YanEtAl2021,OgasawaraIgata2022}. More recently, the
strong-deflection problem at exact extremality was shown to develop a
power-law critical contribution in addition to logarithmic structure
\cite{FeleppaBozzaDeGraaf2026}. Related work on the Kerr photon shell
has emphasized the nontrivial structure of the simultaneous
near-critical and near-extremal regime
\cite{ShirHadar2025}. These results expose a singular transition:
fixed subextremal Kerr exhibits the familiar logarithmic
strong-deflection behavior, whereas the exactly extremal prograde
problem possesses a stronger power-law critical behavior.

The purpose of this work is to resolve the transition between these
two regimes. Rather than taking the near-critical and near-extremal
limits separately, we take them simultaneously and construct a
uniform asymptotic expansion. To our knowledge, a parameter-uniform
analytic strong-deflection formula that explicitly connects the
subextremal logarithmic regime to the extremal power-law regime has
not previously been obtained for prograde equatorial Kerr null
geodesics.

We introduce the near-extremality parameter

\begin{equation}
\kappa
=
\sqrt{
1-\frac{a^2}{M^2}
},
\label{eq:kappa_intro}
\end{equation}

so that \(0<\kappa\ll1\) describes the near-extremal regime and
\(\kappa\to0\) corresponds to exact extremality. Let \(b\) denote the
energy-normalized azimuthal impact parameter of the equatorial null
geodesic, and let \(b_c\) denote its exact critical value for the
prograde equatorial photon orbit. Their precise definitions are given
in Sec.~\ref{sec:geodesics}.

We show that the distinguished joint near-critical and near-extremal
limit occurs when the physical impact-parameter offset from the exact
critical value satisfies

\begin{equation}
b-b_c
=
\Order(M\kappa).
\label{eq:distinguished_intro}
\end{equation}

This scaling motivates the dimensionless crossover coordinate

\begin{equation}
\chi
=
\frac{
b-b_c
}{
M\kappa
}.
\label{eq:chi_intro}
\end{equation}

Holding \(\chi\) fixed while taking \(\kappa\to0\) resolves the
transition region between the subextremal and extremal asymptotic
descriptions.

In this scaling limit, the exact equatorial Kerr radial potential
reduces at leading order to a quadratic normal form. The corresponding
leading accumulated azimuthal phase, denoted by \(\Phi\), can be
evaluated analytically and is

\begin{equation}
\Phi_{\rm LO}
=
\frac{1}{\kappa}
\ln
\left(
1+
\frac{
2\sqrt{3}
}{
\chi
}
\right).
\label{eq:Phi_intro}
\end{equation}

Equation~\eqref{eq:Phi_intro} is the leading uniform crossover law of
this work. It contains both endpoint behaviors within a single scaled
description. On the near-critical side of the crossover,
\(\chi\ll1\),

\begin{equation}
\Phi_{\rm LO}
\sim
\frac{1}{\kappa}
\left[
-\ln\chi+\ln(2\sqrt{3})
\right],
\label{eq:Phi_small_intro}
\end{equation}

recovering the near-extremal form of the logarithmic
strong-deflection singularity. In the opposite extremal-like overlap,

\begin{equation}
1
\ll
\chi
\ll
\kappa^{-1},
\label{eq:chi_extremal_intro}
\end{equation}

the same expression yields

\begin{equation}
\Phi_{\rm LO}
\sim
\frac{
2\sqrt{3}M
}{
b-b_c
},
\label{eq:Phi_large_intro}
\end{equation}

which reproduces the leading power-law behavior characteristic of the
exactly extremal prograde critical orbit. The logarithmic and
power-law regimes therefore arise as opposite asymptotic limits of a
single near-extremal crossover rather than as disconnected
strong-deflection descriptions.

We next carry the matched asymptotic expansion to
next-to-leading order. After matching the near-horizon and outer Kerr
regions and recentering the expansion on the exact physical critical
impact parameter, we obtain a subleading correction containing a
logarithmic term and a finite matching constant. We test the resulting
expression directly against high-precision integrations of the exact
Kerr geodesic equations. Across a survey of 55 independently computed
trajectories, the next-to-leading-order approximation improves on the
leading uniform result in every tested case.

The same crossover reorganizes the hierarchy of higher-order images.
For subextremal black holes, near-critical photon trajectories generate
a characteristic exponential hierarchy of relativistic images and
photon-ring substructure
\cite{Bozza2002,Bozza2003,BozzaEtAl2005,
GrallaHolzWald2019,GrallaLupsasca2020Lensing,
GrallaLupsascaMarrone2020,HadarEtAl2021}.
More recent work has further emphasized the critical dynamics,
scaling structure, and observable properties of successive
photon-ring images
\cite{HadarKapecLupsascaStrominger2022,
KocherlakotaEtAl2024,WaliaEtAl2025,
BerensEtAl2026,YanLiYue2026}.

In the near-extremal problem considered here, we obtain a closed
uniform image law that interpolates between algebraic inverse-order
accumulation and the usual exponential hierarchy. If \(n\) denotes
the image order, the crossover is governed at large \(n\) by the
combined variable \(\kappa n\). The transition occurs for
\(\kappa n=\Order(1)\), corresponding to a characteristic crossover
image order that grows as \(\kappa^{-1}\).

Finally, we extend the analysis to coordinate time. Strong-field
black-hole lensing time delays have previously been investigated for
both static and rotating geometries
\cite{BozzaMancini2004,HsiehLeeLin2021}.
More recent work has emphasized the role of image-order-dependent
arrival times and photon-ring delay scales in strong-field imaging
\cite{KocherlakotaEtAl2024,WaliaEtAl2025,
RojasPaterninaCardenasAvendano2026,YanLiYue2026}.
Here we derive the corresponding uniform near-extremal
image-to-image coordinate-time-delay law and validate it using
independently integrated exact Kerr arrival times. Because the
additive large-distance regularization constant cancels between image
trajectories, the delay differences provide an independent test of
the same near-extremal asymptotic structure.

The remainder of this paper is organized as follows.
Section~\ref{sec:geodesics} reviews the exact equatorial Kerr
null-geodesic equations and the prograde critical orbit.
Section~\ref{sec:normalform} derives the joint near-extremal and
near-critical radial normal form.
Section~\ref{sec:leading} obtains the leading uniform
strong-deflection crossover.
Section~\ref{sec:nlo} develops the matched next-to-leading-order
expansion.
Section~\ref{sec:images} derives the higher-order image crossover,
and Sec.~\ref{sec:time} develops the coordinate-time-delay law.
Section~\ref{sec:numerics} presents the direct validation against
exact Kerr geodesics.
The asymptotic interpretation, limitations, and possible extensions
are discussed in Sec.~\ref{sec:discussion}, and the main results are
summarized in Sec.~\ref{sec:conclusions}.


\section{Exact Equatorial Kerr Null Geodesics}
\label{sec:geodesics}

\subsection{Near-extremal Kerr geometry}

We work in Boyer--Lindquist coordinates
\((t,r,\theta,\phi)\) and adopt the standard Kerr conventions
\cite{Carter1968,BardeenPressTeukolsky1972,
Teo2003,GrallaLupsasca2020Null}. The metric functions relevant to the
equatorial null-geodesic problem are

\begin{equation}
\Sigma
=
r^2+a^2\cos^2\theta,
\qquad
\Delta
=
r^2-2Mr+a^2.
\label{eq:Sigma_Delta}
\end{equation}

The roots of \(\Delta=0\) determine the outer and inner horizon
radii,

\begin{equation}
r_{\pm}
=
M
\pm
\sqrt{M^2-a^2},
\label{eq:horizons}
\end{equation}

where \(r_+\) is the outer event horizon and \(r_-\) is the inner
horizon.

Throughout this work, we consider the near-extremal prograde branch
with \(a>0\). Using the near-extremality parameter introduced in
Sec.~\ref{sec:introduction},

\begin{equation}
\kappa
=
\sqrt{
1-\frac{a^2}{M^2}
},
\qquad
\frac{a}{M}
=
\sqrt{1-\kappa^2},
\label{eq:kappa_geometry}
\end{equation}

the horizon radii become

\begin{equation}
r_{\pm}
=
M(1\pm\kappa).
\label{eq:horizons_kappa}
\end{equation}

The near-extremal regime corresponds to \(0<\kappa\ll1\), with exact
extremality obtained as \(\kappa\to0\). The horizon separation is

\begin{equation}
r_+-r_-
=
2M\kappa,
\label{eq:horizon_separation}
\end{equation}

showing that \(M\kappa\) sets the natural near-horizon radial scale.
As shown below, the prograde critical photon orbit approaches the
outer horizon on this same scale, providing the geometric origin of
the joint near-extremal and near-critical scaling developed in the
subsequent sections.

\subsection{Equatorial null geodesics}

We restrict to equatorial null geodesics,
\(\theta=\pi/2\), for which the Carter constant vanishes.
Let \(p_\mu\) denote the covariant photon momentum. Stationarity and
axial symmetry imply the conserved energy and axial angular momentum

\begin{equation}
E
=
-p_t,
\qquad
L_z
=
p_\phi,
\end{equation}

with \(E>0\) for the future-directed scattering trajectories considered
here. The remaining geodesic equations follow from the standard
separability of Kerr spacetime
\cite{Carter1968,Wilkins1972,GrallaLupsasca2020Null,
CompereLiuLong2022}.

It is convenient to introduce the energy-normalized azimuthal impact
parameter

\begin{equation}
b
=
\frac{L_z}{E},
\label{eq:b_def}
\end{equation}

together with the radial combination

\begin{equation}
P(r)
=
r^2+a^2-ab.
\label{eq:P_def}
\end{equation}

After dividing the standard Kerr radial potential by \(E^2\), the
equatorial null radial potential becomes

\begin{equation}
R(r)
=
P(r)^2
-
\Delta(b-a)^2.
\label{eq:R_exact}
\end{equation}

Radial turning points satisfy

\begin{equation}
R(r)
=
0.
\end{equation}

On the outgoing branch, the separated radial and azimuthal equations
combine to give

\begin{equation}
\frac{\dd\phi}{\dd r}
=
\frac{
(b-a)+aP/\Delta
}{
\sqrt{R}
}.
\label{eq:dphi_dr}
\end{equation}

For a scattering trajectory arriving from infinity, reaching an outer
radial turning point \(r_0\), and returning to infinity, the ingoing
and outgoing radial legs give equal contributions to the accumulated
azimuthal phase. The total two-leg azimuthal accumulation is therefore

\begin{equation}
\Phi(b,a)
=
2
\int_{r_0}^{\infty}
\frac{
(b-a)+aP/\Delta
}{
\sqrt{R}
}
\,\dd r.
\label{eq:Phi_exact}
\end{equation}

The corresponding bending angle is

\begin{equation}
\hat{\alpha}
=
\Phi-\pi.
\label{eq:bending_angle}
\end{equation}

In what follows, we work primarily with \(\Phi\), since its divergent
part directly encodes the strong-deflection structure.

\subsection{Prograde critical orbit}

The unstable equatorial circular photon orbit is characterized by a
double root of the radial potential,

\begin{equation}
R(r_c)
=
0,
\qquad
R_{,r}(r_c)
=
0,
\label{eq:critical_conditions}
\end{equation}

where \(r_c\) denotes the critical photon-orbit radius and
\(R_{,r}\equiv\partial R/\partial r\) is evaluated at fixed impact
parameter
\cite{BardeenPressTeukolsky1972,Teo2003,
GrallaLupsasca2020Null,HsiehLeeLin2021Strong}.

For the prograde branch, the exact critical radius is

\begin{equation}
\frac{r_c}{M}
=
2
\left[
1+
\cos
\left(
\frac{2}{3}
\cos^{-1}
\left(
-\frac{a}{M}
\right)
\right)
\right],
\label{eq:rc_exact}
\end{equation}

and the corresponding critical impact parameter is

\begin{equation}
\frac{b_c}{M}
=
\frac{a}{M}
+
\left(
\frac{r_c}{M}
\right)^{3/2}.
\label{eq:bc_exact}
\end{equation}

Expanding these quantities in the near-extremal limit
\(\kappa\ll1\) gives

\begin{align}
\frac{r_c}{M}
&=
1+
\frac{2}{\sqrt{3}}\kappa
+
\Order(\kappa^2),
\nonumber\\
\frac{b_c}{M}
&=
2+
\sqrt{3}\,\kappa
+
\Order(\kappa^2),
\nonumber\\
\frac{r_c-r_+}{M}
&=
\left(
\frac{2}{\sqrt{3}}-1
\right)\kappa
+
\Order(\kappa^2).
\label{eq:critical_near_extremal}
\end{align}

Thus the prograde critical orbit approaches the outer event horizon
on the same \(M\kappa\) radial scale that controls the horizon
separation. The critical impact parameter likewise departs from its
exact-extremal value,

\begin{equation}
b_c
\longrightarrow
2M,
\qquad
\kappa\to0,
\end{equation}

by an amount of order \(M\kappa\). These common scalings signal that
the near-critical and near-extremal limits cannot be treated
independently and motivate the joint scaling introduced in
Sec.~\ref{sec:normalform}. Higher-order expansions of \(r_c\) and
\(b_c\) are collected in Appendix~\ref{app:critical}.


\section{Joint Near-Extremal and Near-Critical Normal Form}
\label{sec:normalform}

\subsection{Near-horizon scaling}

Motivated by the fact that the prograde critical orbit approaches the
outer event horizon on the radial scale \(M\kappa\), we introduce the
dimensionless scaled variables

\begin{equation}
\frac{r}{M}
=
1+\kappa y,
\qquad
\frac{b}{M}
=
2+\kappa X.
\label{eq:scaled_variables}
\end{equation}

Here \(y\) resolves radial distances from the extremal horizon
\(r=M\) on the near-horizon scale \(M\kappa\), while \(X\) resolves
impact-parameter displacements from the extremal critical value
\(b=2M\) on the same near-extremal scale. At this stage, \(X\) is an
intermediate leading-order scaled impact-parameter coordinate; the
physical displacement from the exact critical value \(b_c\) will be
introduced below.

We consider the limit

\begin{equation}
\kappa\to0
\end{equation}

with \(y\) and \(X\) held fixed. Substituting
Eq.~\eqref{eq:scaled_variables} into the exact equatorial Kerr radial
potential and expanding in \(\kappa\) gives

\begin{equation}
\frac{R}{M^4}
=
\kappa^2
\left(
3y^2-4Xy+X^2+1
\right)
+
\Order(\kappa^3).
\label{eq:R_normal_form}
\end{equation}

We therefore define the reduced near-horizon radial potential

\begin{equation}
Q(y;X)
=
3y^2-4Xy+X^2+1,
\label{eq:Q_normal}
\end{equation}

so that

\begin{equation}
\frac{R}{M^4}
=
\kappa^2Q(y;X)
+
\Order(\kappa^3).
\label{eq:R_Q_normal}
\end{equation}

Equation~\eqref{eq:R_normal_form} shows that the leading joint
near-extremal and near-critical radial dynamics are governed by a
quadratic normal form. The intermediate expansions of
\(\Delta\), \(P(r)\), and the exact radial potential are collected in
Appendix~\ref{app:normal_form}.

\subsection{Reduced critical point}

The two roots of the reduced potential,

\begin{equation}
Q(y;X)=0,
\end{equation}

are

\begin{equation}
y_{\pm}
=
\frac{
2X\pm\sqrt{X^2-3}
}{
3
}.
\label{eq:y_pm}
\end{equation}

The two roots coalesce when the discriminant vanishes,

\begin{equation}
X^2-3
=
0.
\end{equation}

For the prograde near-critical branch considered here, the relevant
solution is

\begin{equation}
X_c
=
\sqrt{3},
\qquad
y_c
=
\frac{2}{\sqrt{3}}.
\label{eq:reduced_critical_point}
\end{equation}

Through the scalings in Eq.~\eqref{eq:scaled_variables}, these values
reproduce the leading near-extremal critical impact parameter and
critical photon-orbit radius of the exact Kerr problem,

\begin{align}
\frac{b_c}{M}
&=
2+\sqrt{3}\,\kappa+\Order(\kappa^2),
\nonumber\\
\frac{r_c}{M}
&=
1+\frac{2}{\sqrt{3}}\kappa+\Order(\kappa^2).
\label{eq:critical_recovered_normal}
\end{align}

The reduced normal form therefore preserves the critical double-root
structure of the exact Kerr radial potential in the joint
near-extremal and near-critical limit.

\subsection{Distinguished impact-parameter scaling}

Near-extremal Kerr geodesics exhibit a characteristic near-horizon
scaling structure that has been explored in studies of high-spin
photon emission, critical geodesics, and near-horizon motion
\cite{PorfyriadisShiStrominger2017,
GrallaLupsascaStrominger2018,
LupsascaPorfyriadisShi2018,
KapecLupsasca2020,
YanEtAl2021,
OgasawaraIgata2022}.
Related work on near-critical Kerr dynamics has also identified
nontrivial structure when the critical and near-extremal limits are
considered simultaneously
\cite{GrallaLupsasca2020Null,ShirHadar2025}.

The reduced critical point occurs at \(X_c=\sqrt{3}\). Since the
scaled impact parameter satisfies

\begin{equation}
\frac{b}{M}
=
2+\kappa X,
\end{equation}

while the exact critical impact parameter has the near-extremal
expansion

\begin{equation}
\frac{b_c}{M}
=
2+\sqrt{3}\,\kappa+\Order(\kappa^2),
\end{equation}

their physical difference is

\begin{equation}
\frac{b-b_c}{M}
=
\kappa
\left(
X-\sqrt{3}
\right)
+
\Order(\kappa^2).
\label{eq:b_offset_scaled}
\end{equation}

Thus a nontrivial joint near-critical and near-extremal limit is
obtained when the displacement from the critical impact parameter is
of the same order as the near-extremal scale,

\begin{equation}
b-b_c
=
\Order(M\kappa).
\label{eq:distinguished_scaling}
\end{equation}

This motivates the physical dimensionless crossover coordinate

\begin{equation}
\chi
=
\frac{
b-b_c
}{
M\kappa
}.
\label{eq:chi_main}
\end{equation}

The joint limit is defined by

\begin{equation}
\kappa\to0,
\qquad
\chi=\Order(1),
\label{eq:joint_limit_chi}
\end{equation}

with \(\chi\) held fixed. Equation~\eqref{eq:b_offset_scaled} then
implies

\begin{equation}
X
=
\sqrt{3}+\chi+\Order(\kappa).
\label{eq:X_chi_relation}
\end{equation}

At leading order,

\begin{equation}
X
=
\sqrt{3}+\chi.
\label{eq:X_chi_leading}
\end{equation}

Thus \(X-\sqrt{3}\) is the leading reduced displacement from the
critical point, while \(\chi\) measures the physical impact-parameter
offset from the exact Kerr critical value \(b_c\). The subsequent
uniform asymptotic analysis is organized around this distinguished
physical crossover variable.


\section{Leading Uniform Strong-Deflection Law}
\label{sec:leading}

Substituting the joint near-extremal and near-critical scaling of
Sec.~\ref{sec:normalform} into the exact azimuthal integral isolates
the near-horizon contribution responsible for the leading
strong-deflection singularity. In the limit

\begin{equation}
\kappa\to0,
\qquad
X=\Order(1),
\end{equation}

with \(X\) held fixed, the leading two-leg azimuthal accumulation takes
the form

\begin{equation}
\Phi_{\rm LO}
=
\frac{1}{\kappa}F_X(X),
\label{eq:Phi_F_X}
\end{equation}

where the dimensionless reduced crossover function is obtained by
explicit evaluation of the leading near-horizon integral. As shown in
Appendix~\ref{app:F},

\begin{equation}
F_X(X)
=
\ln
\left(
\frac{
X+\sqrt{3}
}{
X-\sqrt{3}
}
\right),
\qquad
X>\sqrt{3}.
\label{eq:F_X}
\end{equation}

The condition \(X>\sqrt{3}\) corresponds to the scattering side of the
reduced critical point, for which the outer radial turning point lies
outside the critical orbit.

Using the leading relation

\begin{equation}
X
=
\sqrt{3}+\chi,
\end{equation}

we define the same crossover function in terms of the physical
coordinate \(\chi\) by

\begin{equation}
F(\chi)
\equiv
F_X(\sqrt{3}+\chi)
=
\ln
\left(
1+\frac{
2\sqrt{3}
}{
\chi
}
\right),
\qquad
\chi>0.
\label{eq:F_chi}
\end{equation}

The leading uniform phase law can therefore be written as

\begin{equation}
\Phi_{\rm LO}
=
\frac{F(\chi)}{\kappa}
=
\frac{1}{\kappa}
\ln
\left(
1+\frac{
2\sqrt{3}
}{
\chi
}
\right).
\label{eq:Phi_LO}
\end{equation}

Equation~\eqref{eq:Phi_LO} is the leading uniform
strong-deflection law of this work. Unlike the conventional
fixed-subextremal strong-deflection expansion, whose critical
divergence is logarithmic
\cite{Bozza2003,BozzaEtAl2005,BozzaEtAl2006,
IyerHansen2009,HsiehLeeLin2021Strong},
Eq.~\eqref{eq:Phi_LO} remains nontrivial when the near-critical and
near-extremal limits are taken simultaneously. Its two asymptotic
limits recover the near-extremal form of the ordinary subextremal
logarithmic behavior and the leading power-law structure of the
exactly extremal prograde problem
\cite{FeleppaBozzaDeGraaf2026}.

\subsection{Logarithmic and power-law limits}

The uniform law in Eq.~\eqref{eq:Phi_LO} connects the two distinct
critical behaviors associated with the subextremal and extremal
endpoints.

On the near-critical side of the crossover,

\begin{equation}
\chi
\ll
1,
\end{equation}

the physical crossover function defined in
Eq.~\eqref{eq:F_chi} has the expansion

\begin{equation}
F(\chi)
=
-\ln\chi
+
\ln(2\sqrt{3})
+
\Order(\chi).
\label{eq:small_chi}
\end{equation}

Hence

\begin{equation}
\Phi_{\rm LO}
=
\frac{1}{\kappa}
\left[
-\ln\chi
+
\ln(2\sqrt{3})
+
\Order(\chi)
\right],
\end{equation}

which reproduces the near-extremal form of the standard logarithmic
strong-deflection singularity of subextremal Kerr
\cite{Bozza2003,BozzaEtAl2005,BozzaEtAl2006,
IyerHansen2009,HsiehLeeLin2021Strong}.

This statement concerns the leading critical behavior in the overlap
with the near-extremal regime. Recovery of the complete
fixed-subextremal next-to-leading-order expansion arbitrarily deep in
the nested limit \(\chi\to0\) requires the additional boundary-layer
analysis discussed in Sec.~\ref{sec:nlo}.

On the opposite side of the crossover, the extremal-like overlap is

\begin{equation}
1
\ll
\chi
\ll
\kappa^{-1}.
\label{eq:extremal_like_chi_domain}
\end{equation}

For \(\chi\gg1\),

\begin{equation}
F(\chi)
=
\frac{
2\sqrt{3}
}{
\chi
}
+
\Order(\chi^{-2}),
\label{eq:F_large_chi}
\end{equation}

so that

\begin{equation}
\Phi_{\rm LO}
\sim
\frac{
2\sqrt{3}
}{
\kappa\chi
}
=
\frac{
2\sqrt{3}M
}{
b-b_c
}.
\label{eq:power_endpoint}
\end{equation}

This reproduces the leading power-law critical behavior of the
prograde exactly extremal strong-deflection problem
\cite{FeleppaBozzaDeGraaf2026}. Since

\begin{equation}
b_c
\longrightarrow
2M
\qquad
(\kappa\to0),
\end{equation}

the impact-parameter offset in this overlap approaches that measured
from the exactly extremal critical value, and the uniform expression
reduces to the corresponding extremal scaling.

The logarithmic near-critical behavior and the extremal power-law
behavior therefore arise as opposite asymptotic limits of the same
near-extremal crossover. The transition is controlled by the ratio of
the physical impact-parameter offset to the near-extremal scale
\(M\kappa\), namely by the crossover coordinate \(\chi\), rather than
by either limiting procedure taken in isolation.


\section{Matched Next-to-Leading-Order Expansion}
\label{sec:nlo}

\subsection{Turning-point expansion}

To organize the next-to-leading-order correction, we parametrize the
outer radial turning point as

\begin{equation}
\frac{r_0}{M}
=
1+\kappa y_0,
\label{eq:r0_scaled_nlo}
\end{equation}

where \(y_0\) is the scaled turning-point coordinate introduced by the
near-horizon radial scaling of Sec.~\ref{sec:normalform}.

The exact turning-point condition \(R(r_0)=0\) can be solved on the
physical prograde scattering branch as

\begin{equation}
b
=
a
+
\frac{
r_0^2
}{
a+\sqrt{\Delta_0}
},
\qquad
\Delta_0
\equiv
\Delta(r_0)
=
r_0^2-2Mr_0+a^2.
\label{eq:turning_exact}
\end{equation}

Expanding this exact relation in the joint near-extremal and
near-critical scaling gives

\begin{align}
\frac{b}{M}
&=
2+\kappa X+\kappa^2\beta(X)
+\Order(\kappa^3),
\label{eq:b_turning_series}
\\
\beta(X)
&=
\frac{
2X^2
-
2X\sqrt{X^2-3}
-
3
}{
9
}.
\label{eq:beta_X}
\end{align}

Here \(X\) is the leading scaled impact-parameter coordinate defined
in Sec.~\ref{sec:normalform}, while \(\beta(X)\) is the corresponding
subleading turning-point coefficient.

At the reduced critical point \(X=X_c=\sqrt{3}\),

\begin{equation}
\beta_c
\equiv
\beta(\sqrt{3})
=
\frac{1}{3},
\label{eq:beta_c}
\end{equation}

in agreement with the \(\Order(\kappa^2)\) term in the near-extremal
expansion of the exact critical impact parameter. The intermediate
relations among \(y_0\), \(X\), and the turning-point root are
collected in Appendix~\ref{app:nlo}.

\subsection{Matched next-to-leading-order correction}

The next-to-leading-order expansion requires matching the
near-horizon contribution to the outer Kerr region while keeping the
physical turning point fixed. The detailed expansions of the radial
potential and azimuthal integrand, together with the cancellation of
the intermediate matching scale, are given in
Appendix~\ref{app:nlo}.

At this order it is important to distinguish the intermediate
leading-order coordinate \(X\) from the physical coordinate centered
on the exact critical impact parameter. We therefore define

\begin{equation}
x
\equiv
\sqrt{3}+\chi,
\label{eq:x_chi_nlo}
\end{equation}

where \(\chi=(b-b_c)/(M\kappa)\) is the physical crossover coordinate
introduced in Eq.~\eqref{eq:chi_main}. The relation between \(X\) and
\(x\) is

\begin{equation}
X
=
x
+
\Order(\kappa),
\label{eq:X_x_nlo}
\end{equation}

with the explicit subleading recentering derived in
Appendix~\ref{app:nlo}.

It is also convenient to define

\begin{equation}
D_x
\equiv
\sqrt{x^2-3},
\qquad
x>\sqrt{3}.
\label{eq:Dx_nlo}
\end{equation}

After near--outer matching and recentering on the exact physical
critical impact parameter, the accumulated azimuthal phase takes the
form

\begin{align}
\Phi_{\rm NLO}
={}&
\frac{1}{\kappa}
\ln
\left(
\frac{x+\sqrt{3}}{x-\sqrt{3}}
\right)
\nonumber\\
&-
\frac{4\sqrt{3}}{9}
\ln
\left(
\kappa D_x
\right)
+
C_{\phi}
+
o(1).
\label{eq:Phi_NLO}
\end{align}

The finite matching constant is

\begin{equation}
C_{\phi}
=
\frac{4\sqrt{3}}{9}
\left[
3+\ln\!\left(36(2-\sqrt{3})\right)
\right]
-\frac{4}{3}
\simeq
2.7208672467.
\label{eq:Cphi_exact}
\end{equation}

Equation~\eqref{eq:Phi_NLO} is the matched next-to-leading-order
uniform phase law. It is tested directly against exact Kerr geodesics
in Sec.~\ref{sec:numerics}.

\subsection{Extremal-like overlap and domain of validity}

To connect the matched near-extremal expansion with the exactly
extremal endpoint, we introduce the dimensionless impact-parameter
offset from the extremal critical value \(2M\),

\begin{equation}
\delta_b
\equiv
\frac{
b-2M
}{
M
}.
\label{eq:delta_b_extremal}
\end{equation}

In the extremal-like overlap,

\begin{equation}
\kappa
\ll
\delta_b
\ll
1,
\label{eq:delta_b_extremal_domain}
\end{equation}

equivalently

\begin{equation}
1
\ll
\chi
\ll
\kappa^{-1},
\end{equation}

Eq.~\eqref{eq:Phi_NLO} reduces to

\begin{equation}
\Phi
\sim
\frac{
2\sqrt{3}
}{
\delta_b
}
-
\frac{
4\sqrt{3}
}{
9
}
\ln\delta_b
+
C_{\phi}
+
\cdots.
\label{eq:extremal_overlap}
\end{equation}

This reproduces the characteristic power-law plus logarithmic
structure of the prograde exactly extremal strong-deflection limit
\cite{FeleppaBozzaDeGraaf2026}. A direct coefficient-by-coefficient
comparison with the exact-extremal expansion is given in
Appendix~\ref{app:nlo}.

The matched expansion in Eq.~\eqref{eq:Phi_NLO} is uniform for fixed

\begin{equation}
\chi>0
\end{equation}

as \(\kappa\to0\). It is not intended to reproduce the complete
fixed-subextremal next-to-leading-order coefficients arbitrarily deep
in the nested limit

\begin{equation}
\chi\to0.
\end{equation}

Resolving that endpoint requires an additional fixed-subextremal
boundary-layer analysis matched to the present joint expansion.
This limitation affects the detailed subleading coefficients but does
not alter the leading crossover law or the fixed-\(\chi\) matched
correction derived here.


\section{Higher-Order Image Crossover}
\label{sec:images}

Higher-order image branches arise from near-critical trajectories that
execute additional windings around the black hole before escaping.
Within the present equatorial scattering problem, we characterize the
\(n\)th branch by the phase condition

\begin{equation}
\Phi(b_n,a)
=
\Lcal_n,
\qquad
\Lcal_n
=
2\pi n+\psi,
\label{eq:image_phase_condition}
\end{equation}

where \(n\in\mathbb{N}\) is the discrete image-order or winding
label, \(b_n\) is the corresponding impact parameter, and
\(\psi=\Order(1)\) is a residual phase determined by the
source--observer geometry and by the convention used for the global
lens map. For a fixed asymptotic configuration, \(\psi\) is the
azimuthal offset remaining after the integer winding contribution
\(2\pi n\) is removed. We retain it explicitly to separate this global
geometric information from the local near-critical dynamics.

For the \(n\)th branch, define

\begin{equation}
\chi_n
=
\frac{
b_n-b_c
}{
M\kappa
}.
\label{eq:chi_n}
\end{equation}

Substituting Eq.~\eqref{eq:image_phase_condition} into the leading
uniform phase law Eq.~\eqref{eq:Phi_LO} gives

\begin{equation}
\ln
\left(
1+
\frac{
2\sqrt{3}
}{
\chi_n
}
\right)
=
\kappa\Lcal_n.
\end{equation}

Solving for \(\chi_n\) and using Eq.~\eqref{eq:chi_n} yields

\begin{equation}
\frac{
b_n-b_c
}{
M
}
=
\frac{
2\sqrt{3}\,\kappa
}{
\exp(\kappa\Lcal_n)-1
}.
\label{eq:image_law}
\end{equation}

Equation~\eqref{eq:image_law} is the leading uniform law for the
higher-order impact parameters. Since its derivation assumes
near-critical strong deflection, the accumulated phase must satisfy

\begin{equation}
\Lcal_n
\gg
1.
\label{eq:image_strong_deflection_condition}
\end{equation}

The transition between the two limiting accumulation laws is governed
by the combination \(\kappa\Lcal_n\).

In the regime

\begin{equation}
1
\ll
\Lcal_n
\ll
\kappa^{-1},
\label{eq:image_algebraic_domain}
\end{equation}

one has \(\kappa\Lcal_n\ll1\), and
Eq.~\eqref{eq:image_law} reduces to

\begin{equation}
\frac{
b_n-b_c
}{
M
}
\sim
\frac{
2\sqrt{3}
}{
\Lcal_n
}.
\label{eq:image_algebraic_limit}
\end{equation}

The approach to the critical impact parameter is therefore algebraic.
The condition \(\Lcal_n\gg1\) is essential: it ensures that

\begin{equation}
\frac{
b_n-b_c
}{
M
}
\ll
1,
\end{equation}

so that the corresponding trajectory remains parametrically close to
the critical orbit.

For the reference choice \(\psi=0\),

\begin{equation}
\Lcal_n
=
2\pi n,
\end{equation}

and Eq.~\eqref{eq:image_algebraic_limit} becomes

\begin{equation}
\frac{
b_n-b_c
}{
M
}
\sim
\frac{
\sqrt{3}
}{
\pi n
},
\qquad
1
\ll
n
\ll
\kappa^{-1}.
\label{eq:image_inverse_n}
\end{equation}

At larger \(\kappa\Lcal_n\), Eq.~\eqref{eq:image_law} instead gives

\begin{equation}
\frac{
b_n-b_c
}{
M
}
\sim
2\sqrt{3}\,\kappa
e^{-\kappa\Lcal_n},
\qquad
\kappa\Lcal_n\gg1,
\label{eq:image_exponential_limit}
\end{equation}

recovering the exponential hierarchy characteristic of ordinary
subextremal strong deflection and photon-ring substructure
\cite{Bozza2002,Bozza2003,BozzaEtAl2005,
GrallaHolzWald2019,GrallaLupsasca2020Lensing,
GrallaLupsascaMarrone2020,HadarEtAl2021,
HadarKapecLupsascaStrominger2022,BerensEtAl2026}.

The change between these behaviors occurs when

\begin{equation}
\kappa\Lcal_n
=
\Order(1).
\label{eq:image_crossover_condition}
\end{equation}

Since \(\psi=\Order(1)\), it contributes only subleading corrections
to the large-\(n\) scaling, giving the characteristic image order

\begin{equation}
n_{\rm cross}
\sim
\frac{
1
}{
2\pi\kappa
}.
\label{eq:ncross}
\end{equation}

Hence the number of orders exhibiting the algebraic hierarchy grows
parametrically as \(\kappa^{-1}\) in the near-extremal limit.

This structure is especially simple in the joint high-order and
near-extremal limit. For \(\psi=0\), define

\begin{equation}
s
\equiv
2\pi\kappa n
=
\kappa\Lcal_n.
\label{eq:s_image}
\end{equation}

We consider

\begin{equation}
\kappa\to0,
\qquad
n\to\infty,
\qquad
s=\Order(1),
\label{eq:image_double_scaling}
\end{equation}

with \(s\) held fixed. Because \(n\) is discrete, this limit is
understood along sequences of integer image orders and corresponding
near-extremality parameters satisfying
\(2\pi\kappa n\to s\).

Equation~\eqref{eq:image_law} then becomes

\begin{equation}
n
\frac{
b_n-b_c
}{
M
}
=
\frac{
\sqrt{3}
}{
\pi
}
\frac{
s
}{
e^s-1
}.
\label{eq:image_master}
\end{equation}

This parameter-free relation provides the master curve for the
equatorial Kerr image sequence. For

\begin{equation}
s
\ll
1,
\qquad
n
\gg
1,
\end{equation}

it approaches

\begin{equation}
n
\frac{
b_n-b_c
}{
M
}
\longrightarrow
\frac{
\sqrt{3}
}{
\pi
},
\qquad
s\to0,
\label{eq:image_master_small_s}
\end{equation}

which reproduces the inverse-order law. For \(s\gg1\),

\begin{equation}
n
\frac{
b_n-b_c
}{
M
}
\sim
\frac{
\sqrt{3}
}{
\pi
}
s e^{-s}.
\label{eq:image_master_large_s}
\end{equation}

Thus Eq.~\eqref{eq:image_master} continuously connects algebraic
accumulation at small \(s\) to exponential suppression at large
\(s\). It also shows explicitly that neither \(n\) nor \(\kappa\)
alone determines the transition; the relevant variable is their
product.

The fixed-\(s=\Order(1)\) double-scaling limit is the directly
controlled regime of the present construction. Formally extending
Eq.~\eqref{eq:image_master} to arbitrarily large \(s\) gives

\begin{equation}
\chi_n
=
\frac{
2\sqrt{3}
}{
e^s-1
}
\longrightarrow
0,
\end{equation}

which enters the nested fixed-subextremal boundary layer discussed
below.

The next-to-leading-order correction to \(b_n\) follows by
perturbatively inverting the matched phase law
Eq.~\eqref{eq:Phi_NLO}. It introduces no new scale, and its explicit
form is given in Appendix~\ref{app:nlo}. The expansion is controlled
for fixed positive \(\chi_n\), equivalently fixed finite
\(\kappa\Lcal_n>0\), as \(\kappa\to0\). Sending
\(\kappa\Lcal_n\) arbitrarily far into the large-phase regime instead
drives \(\chi_n\to0\), where the additional boundary-layer analysis
of Sec.~\ref{sec:nlo} is required.

The prediction of Eq.~\eqref{eq:image_master} is tested numerically
against exact Kerr phase solutions in Sec.~\ref{sec:numerics}.


\section{Uniform Coordinate-Time Delays}
\label{sec:time}

Strong-field image delays provide an independent probe of the same
near-critical geodesics that control the azimuthal strong-deflection
law
\cite{BozzaMancini2004,HsiehLeeLin2021,
WaliaEtAl2025,RojasPaterninaCardenasAvendano2026,
YanLiYue2026}.
The absolute Kerr scattering coordinate time diverges at spatial
infinity, so we remove the universal large-radius contribution before
matching the near-horizon and outer regions. We denote the resulting
regularized scattering time by \(T_{\rm reg}\). Different subtraction
conventions shift \(T_{\rm reg}\) by a finite additive constant,
whereas image-to-image delay differences are independent of that
convention.

In the joint near-extremal and near-critical scaling, the leading
near-horizon temporal and azimuthal contributions satisfy

\begin{equation}
\dd t
\simeq
2M\,\dd\phi,
\label{eq:dt_dphi_leading}
\end{equation}

consistent with the Kerr horizon angular velocity

\begin{equation}
\Omega_H
=
\frac{a}{r_+^2+a^2}
=
\frac{a}{2Mr_+}
\longrightarrow
\frac{1}{2M},
\qquad
\kappa\to0.
\label{eq:OmegaH_extremal}
\end{equation}

After matching the near-horizon contribution to the outer Kerr region,
the regularized coordinate time takes the form

\begin{equation}
\frac{T_{\rm reg}}{M}
=
2\Phi
-
2\sqrt{3}
\ln
\left(
\kappa D_x
\right)
+
C_{\Delta}
+
o(1),
\label{eq:Treg_phi}
\end{equation}

where

\begin{equation}
x
=
\sqrt{3}+\chi,
\qquad
D_x
\equiv
\sqrt{x^2-3},
\label{eq:x_Dx_time}
\end{equation}

and \(\Phi\) denotes the accumulated azimuthal phase in the same joint
asymptotic expansion. The finite constant \(C_{\Delta}\) depends on
the adopted large-radius subtraction convention. The explicit
regularization and near--outer matching leading to
Eq.~\eqref{eq:Treg_phi} are given in Appendix~\ref{app:time}.

We now evaluate Eq.~\eqref{eq:Treg_phi} on the higher-order image
sequence introduced in Sec.~\ref{sec:images}. For the \(n\)th image
branch, define

\begin{equation}
x_n
\equiv
\sqrt{3}+\chi_n.
\label{eq:xn_definition_time}
\end{equation}

Using the leading image law, Eq.~\eqref{eq:image_law}, one obtains

\begin{equation}
x_n
=
\sqrt{3}
\coth
\left(
\frac{
\kappa\Lcal_n
}{
2
}
\right),
\label{eq:xn_time}
\end{equation}

and therefore

\begin{equation}
D_n
\equiv
\sqrt{x_n^2-3}
=
\sqrt{3}
\operatorname{csch}
\left(
\frac{
\kappa\Lcal_n
}{
2
}
\right).
\label{eq:Dn_time}
\end{equation}

Using the image condition

\begin{equation}
\Phi_n
=
\Lcal_n,
\end{equation}

the regularized coordinate arrival time of the \(n\)th image branch,
denoted by \(T_n\), becomes

\begin{equation}
\frac{T_n}{M}
=
2\Lcal_n
-
2\sqrt{3}
\ln
\left[
\sqrt{3}\,\kappa
\operatorname{csch}
\left(
\frac{
\kappa\Lcal_n
}{
2
}
\right)
\right]
+
C_{\Delta}
+
o(1).
\label{eq:Tn_main}
\end{equation}

For two image orders \(n\) and \(m\), we define the signed
image-to-image coordinate-time delay by

\begin{equation}
\Delta T_{nm}
\equiv
T_n-T_m.
\label{eq:DeltaT_definition}
\end{equation}

Subtracting Eq.~\eqref{eq:Tn_main} between the two image branches
eliminates the convention-dependent constant \(C_{\Delta}\) and gives

\begin{align}
\frac{\Delta T_{nm}}{M}
={}&
2(\Lcal_n-\Lcal_m)
\nonumber\\
&+
2\sqrt{3}
\ln
\left[
\frac{
\sinh(\kappa\Lcal_n/2)
}{
\sinh(\kappa\Lcal_m/2)
}
\right]
+
o(1).
\label{eq:DeltaT_nm}
\end{align}

Equation~\eqref{eq:DeltaT_nm} is the uniform image-to-image
coordinate-time-delay law. Because the additive regularization
constant cancels, the delay difference is independent of the
large-radius subtraction convention. This is the quantity compared
directly with exact Kerr time integrations in
Sec.~\ref{sec:numerics}. The result also shows that the same scaled
phase variables \(\kappa\Lcal_n\) and \(\kappa\Lcal_m\) that organize
the image-position crossover control the corresponding arrival-time
hierarchy.

\subsection{Extremal-like high-order limit}

In the extremal-like strong-deflection overlap,

\begin{equation}
1
\ll
\Lcal_m,\Lcal_n
\ll
\kappa^{-1},
\label{eq:time_extremal_domain}
\end{equation}

one has

\begin{equation}
\sinh
\left(
\frac{
\kappa\Lcal_j
}{
2
}
\right)
\simeq
\frac{
\kappa\Lcal_j
}{
2
},
\qquad
j\in\{m,n\}.
\end{equation}

The delay law therefore reduces to

\begin{equation}
\frac{\Delta T_{nm}}{M}
\simeq
2(\Lcal_n-\Lcal_m)
+
2\sqrt{3}
\ln
\left(
\frac{
\Lcal_n
}{
\Lcal_m
}
\right).
\label{eq:DeltaT_extremal_like}
\end{equation}

For neighboring image orders,

\begin{equation}
\Lcal_{n+1}
-
\Lcal_n
=
2\pi,
\end{equation}

so that

\begin{equation}
\frac{\Delta T_{n+1,n}}{M}
\simeq
4\pi
+
2\sqrt{3}
\ln
\left(
\frac{
\Lcal_{n+1}
}{
\Lcal_n
}
\right),
\qquad
1
\ll
\Lcal_n
\ll
\kappa^{-1}.
\label{eq:neighbor_delay_general}
\end{equation}

For the reference phase choice \(\psi=0\), for which

\begin{equation}
\Lcal_n
=
2\pi n,
\end{equation}

Eq.~\eqref{eq:neighbor_delay_general} becomes

\begin{equation}
\frac{\Delta T_{n+1,n}}{M}
\simeq
4\pi
+
2\sqrt{3}
\ln
\left(
\frac{
n+1
}{
n
}
\right).
\label{eq:neighbor_delay}
\end{equation}

Consequently,

\begin{equation}
\frac{\Delta T_{n+1,n}}{M}
\longrightarrow
4\pi,
\qquad
n\to\infty,
\quad
\kappa n\to0,
\label{eq:neighbor_delay_limit}
\end{equation}

or equivalently,

\begin{equation}
\Delta T_{n+1,n}
\longrightarrow
4\pi M,
\label{eq:neighbor_delay_physical}
\end{equation}

within the joint high-order extremal-like regime

\begin{equation}
1
\ll
n
\ll
\kappa^{-1}.
\label{eq:neighbor_delay_domain}
\end{equation}

The limiting spacing has a direct near-horizon interpretation. Since
\(\Omega_H\to(2M)^{-1}\), one additional azimuthal winding of
\(2\pi\) corresponds asymptotically to the coordinate-time increment

\begin{equation}
\frac{
2\pi
}{
\Omega_H
}
\longrightarrow
4\pi M.
\label{eq:horizon_period_limit}
\end{equation}

This limit should not be confused with taking \(n\to\infty\) at fixed
nonzero \(\kappa\). In that ordering of limits, \(\kappa n\)
eventually becomes large and the sequence leaves the extremal-like
overlap. The coordinate-time sector therefore exhibits the same
noncommuting high-order and extremal limits that organize the
image-position hierarchy in Sec.~\ref{sec:images}.


\section{Exact Kerr Numerical Validation}
\label{sec:numerics}

\subsection{Computational strategy}

The asymptotic predictions were tested against quantities computed
directly from the exact equatorial Kerr geodesic equations. For each
value of the near-extremality parameter \(\kappa\), the exact spin was
reconstructed from

\begin{equation}
\frac{a}{M}
=
\sqrt{1-\kappa^2},
\label{eq:a_numeric}
\end{equation}

and the exact critical photon-orbit radius \(r_c\) and critical impact
parameter \(b_c\) were evaluated from
Eqs.~\eqref{eq:rc_exact} and \eqref{eq:bc_exact}.

For a prescribed physical crossover coordinate \(\chi\), the impact
parameter of the benchmark trajectory was set to

\begin{equation}
b
=
b_c
+
M\kappa\chi.
\label{eq:b_numeric}
\end{equation}

The corresponding outer radial turning point \(r_0\) was then obtained
by solving the exact Kerr condition

\begin{equation}
R(r_0)
=
0,
\label{eq:r0_numeric}
\end{equation}

with the physical scattering root selected by

\begin{equation}
r_0
>
r_c
>
r_+.
\label{eq:scattering_root_numeric}
\end{equation}

The exact radial integrands possess an integrable square-root endpoint
singularity at \(r=r_0\). Before numerical quadrature, this singularity
was removed by a local regularizing change of integration variable.
The explicit transformation and additional implementation details are
given in Appendix~\ref{app:numerics}.

The accumulated azimuthal phases were evaluated from the exact
two-leg Kerr integral in Eq.~\eqref{eq:Phi_exact}, without replacing
the radial potential or azimuthal integrand by their near-extremal
expansions. The regularized coordinate times were likewise computed
from the exact Kerr time integral described in
Appendix~\ref{app:time}.

For the higher-order calculations, image or phase roots were obtained
by solving the exact Kerr phase condition. Asymptotic expressions were
used only, when convenient, to provide efficient initial brackets or
starting estimates; they did not enter the evaluation of the final
benchmark quantities themselves. The resulting numerical calculations
therefore provide independent tests of the leading fixed-\(\chi\)
crossover, the matched next-to-leading-order phase correction, and
the corresponding high-order and coordinate-time crossover laws.

\subsection{Leading-order data collapse}

The leading uniform law was tested on 55 independently integrated
exact Kerr trajectories. We used five near-extremality parameters,

\begin{equation}
\kappa
=
0.05,\,
0.02,\,
0.01,\,
0.005,\,
0.002,
\end{equation}

together with 11 values of the physical crossover coordinate spanning

\begin{equation}
0.1
\leq
\chi
\leq
10.
\end{equation}

For each trajectory, we compare the exact scaled phase
\(\kappa\Phi_{\rm exact}\) with the analytic crossover function
\(F(\chi)\) defined in Eq.~\eqref{eq:F_chi}. To quantify the
leading-order data collapse, define the relative residual

\begin{equation}
\epsilon_{\rm coll}
=
\frac{
\left|
\kappa\Phi_{\rm exact}
-
F(\chi)
\right|
}{
|F(\chi)|
}.
\label{eq:data_collapse_residual}
\end{equation}

At fixed \(\chi\), the exact Kerr sequences converge toward the same
analytic crossover curve as \(\kappa\to0\). The corresponding
residuals \(\epsilon_{\rm coll}\) decrease systematically throughout
the resolved crossover range, providing a direct numerical test of
the fixed-\(\chi\) uniform limit.

\begin{figure*}[t]
\centering
\includegraphics[
width=0.82\textwidth
]{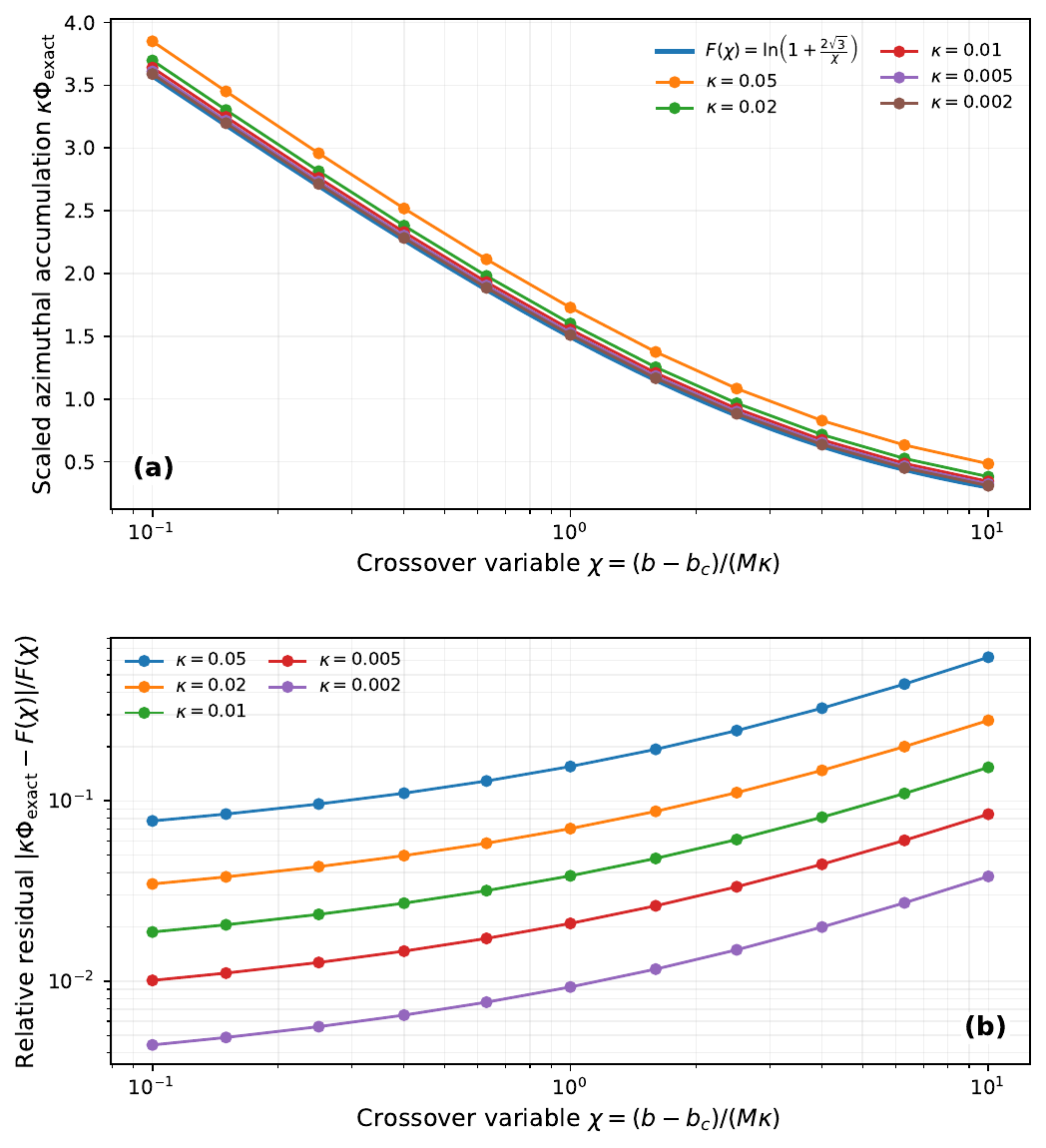}

\caption{
Validation of the leading uniform strong-deflection law.
Panel (a) compares the exact Kerr quantity
\(\kappa\Phi_{\rm exact}\) with the analytic crossover function
\(F(\chi)\) of Eq.~\eqref{eq:F_chi} for five values of the
near-extremality parameter. Panel (b) shows the corresponding relative
residual \(\epsilon_{\rm coll}\) defined in
Eq.~\eqref{eq:data_collapse_residual}. At fixed \(\chi\), the exact
Kerr results converge toward the analytic crossover curve as
\(\kappa\to0\).
}
\label{fig:data_collapse}
\end{figure*}

\subsection{Next-to-leading-order validation}

We quantify the accuracy of the leading- and next-to-leading-order
phase approximations using the relative errors

\begin{equation}
\epsilon_A
=
\frac{
|\Phi_{\rm exact}-\Phi_A|
}{
|\Phi_{\rm exact}|
},
\qquad
A\in\{\mathrm{LO},\mathrm{NLO}\},
\label{eq:numeric_relative_error}
\end{equation}

where \(\Phi_{\rm LO}\) and \(\Phi_{\rm NLO}\) are given by
Eqs.~\eqref{eq:Phi_LO} and \eqref{eq:Phi_NLO}, respectively. We define
the NLO improvement factor by

\begin{equation}
\mathcal I
=
\frac{
\epsilon_{\rm LO}
}{
\epsilon_{\rm NLO}
}.
\label{eq:numeric_improvement}
\end{equation}

\begin{figure*}[t]
\centering
\includegraphics[
width=0.82\textwidth
]{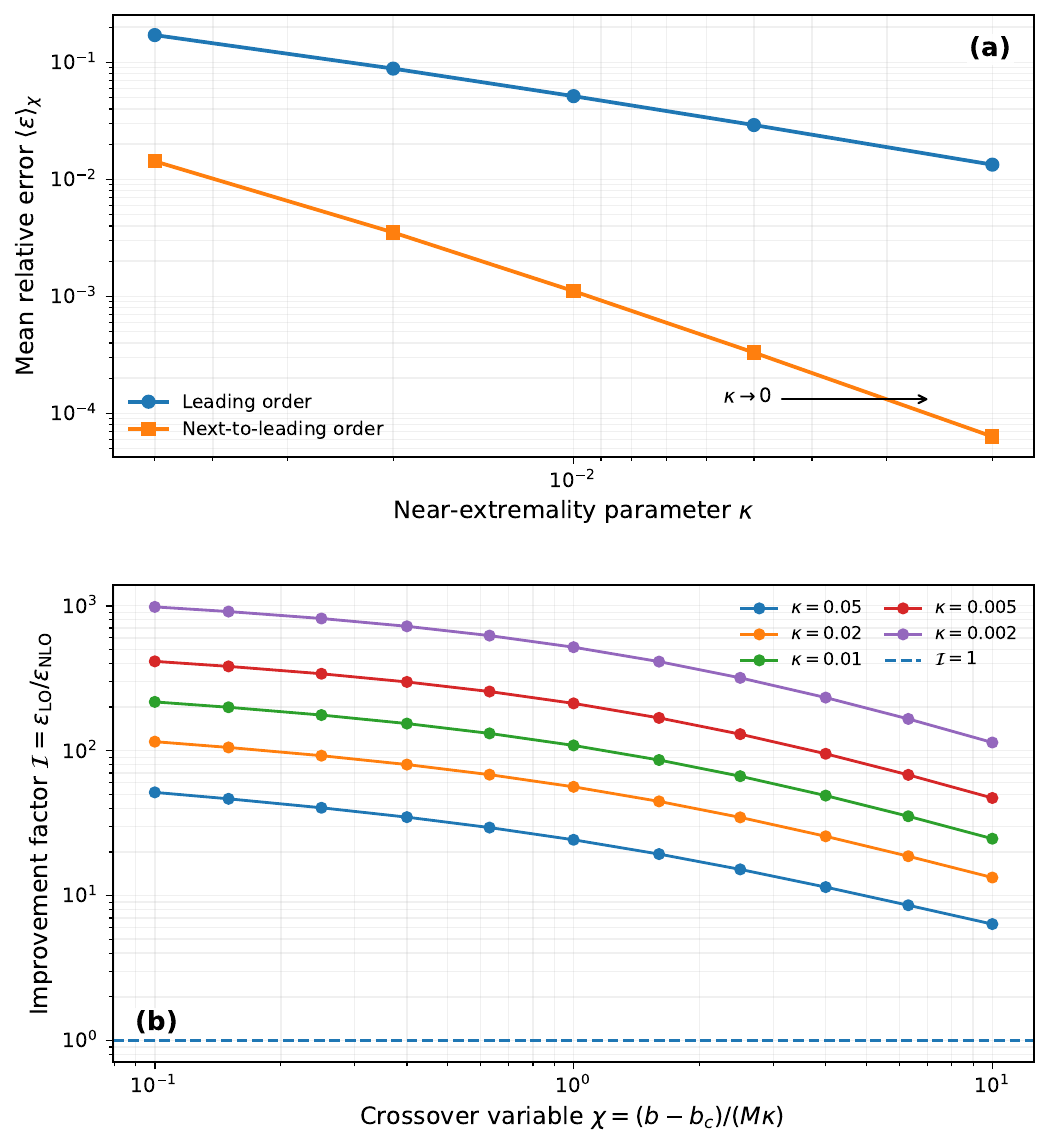}

\caption{
Validation of the matched next-to-leading-order phase expansion.
Panel (a) compares the mean relative errors, averaged over the
11 sampled values of \(\chi\), of the leading- and
next-to-leading-order approximations as functions of \(\kappa\).
Panel (b) shows the improvement factor
\(\mathcal I=\epsilon_{\rm LO}/\epsilon_{\rm NLO}\) throughout the
resolved crossover domain. The dashed line marks
\(\mathcal I=1\); all 55 exact Kerr trajectories lie above this
threshold.
}
\label{fig:nlo_accuracy}
\end{figure*}

Across all 55 exact Kerr trajectories, the NLO approximation improves
on the leading uniform result, with

\begin{equation}
\mathcal I>1
\end{equation}

for every tested case. The improvement factor ranges from
approximately \(6.35\) to \(9.81\times10^2\), with a median value of
\(95.1\). Figure~\ref{fig:nlo_accuracy} shows both the decrease of the
mean relative errors as extremality is approached and the improvement
throughout the resolved crossover domain.

As an independent test of the finite contribution generated by the
near--outer matching, we subtract the analytically predicted singular
terms from the exact Kerr phase and define the effective finite
remainder

\begin{align}
C_{\rm eff}
={}&
\Phi_{\rm exact}
-
\frac{1}{\kappa}
\ln
\left(
\frac{x+\sqrt{3}}{x-\sqrt{3}}
\right)
\nonumber\\
&+
\frac{4\sqrt{3}}{9}
\ln
\left(
\kappa D_x
\right),
\label{eq:Ceff}
\end{align}

where

\begin{equation}
x
=
\sqrt{3}+\chi,
\qquad
D_x
=
\sqrt{x^2-3},
\end{equation}

as defined in Sec.~\ref{sec:nlo}. If the matched expansion is correct,
then at fixed \(\chi>0\),

\begin{equation}
C_{\rm eff}
\longrightarrow
C_\phi,
\qquad
\kappa\to0.
\label{eq:Ceff_limit}
\end{equation}

For each of the 11 fixed values of \(\chi\), the corresponding
sequence \(C_{\rm eff}(\kappa)\) converges toward the analytic matching
constant

\begin{equation}
C_\phi
\simeq
2.7208672467,
\end{equation}

in agreement with Eq.~\eqref{eq:Cphi_exact}. This provides a separate
numerical check of the finite matched contribution in
Eq.~\eqref{eq:Phi_NLO}, beyond the comparison of total phase errors
alone.

\subsection{Exact higher-order image roots}

The uniform image law was tested against higher-order image impact
parameters obtained by independently solving the exact Kerr phase
condition

\begin{equation}
\Phi_{\rm exact}(b_n)
=
2\pi n+\psi.
\label{eq:image_exact_numeric}
\end{equation}

For the numerical validation presented here, we adopt the reference
choice \(\psi=0\), so that each physical image branch satisfies
\(\Phi_{\rm exact}(b_n)=2\pi n\) with
\(n\in\mathbb{N}\).

The test was organized using the scaled image-order variable

\begin{equation}
s
=
2\pi\kappa n,
\label{eq:s_numeric}
\end{equation}

introduced in Sec.~\ref{sec:images}. We used
\(\kappa=0.02,\,0.01,\,0.005\) and selected integer image orders so
that the resulting values of \(s\) lie close to the four target
crossover scales \(0.5,\,1,\,2,\) and \(4\). Specifically, the image
orders were
\(n=\{4,8,16,32\}\) for \(\kappa=0.02\),
\(n=\{8,16,32,64\}\) for \(\kappa=0.01\), and
\(n=\{16,32,64,127\}\) for \(\kappa=0.005\).

For each numerical point, the actual scaled coordinate
\(s=2\pi\kappa n\) was used in the comparison with the analytic master
curve. The resulting sequences are therefore approximately fixed in
\(s\), while retaining the physically required integer image order.
As \(\kappa\) decreases along a target-\(s\) sequence,
\(n=\Order(\kappa^{-1})\), realizing the joint high-order and
near-extremal scaling underlying Eq.~\eqref{eq:image_master}.

The asymptotic image formula was used only to guide numerical
bracketing. The final value of each \(b_n\) was obtained by solving the
exact phase condition. For every one of the 12 roots, the
next-to-leading-order image prediction lies closer to the exact Kerr
solution than the leading approximation. The final root residuals
satisfy

\begin{equation}
\left|
\Phi_{\rm exact}(b_n)-2\pi n
\right|
<
4\times10^{-20},
\label{eq:image_root_residual}
\end{equation}

and are therefore negligible compared with the finite-\(\kappa\)
asymptotic errors.

\begin{figure*}[t]
\centering
\includegraphics[
width=0.82\textwidth
]{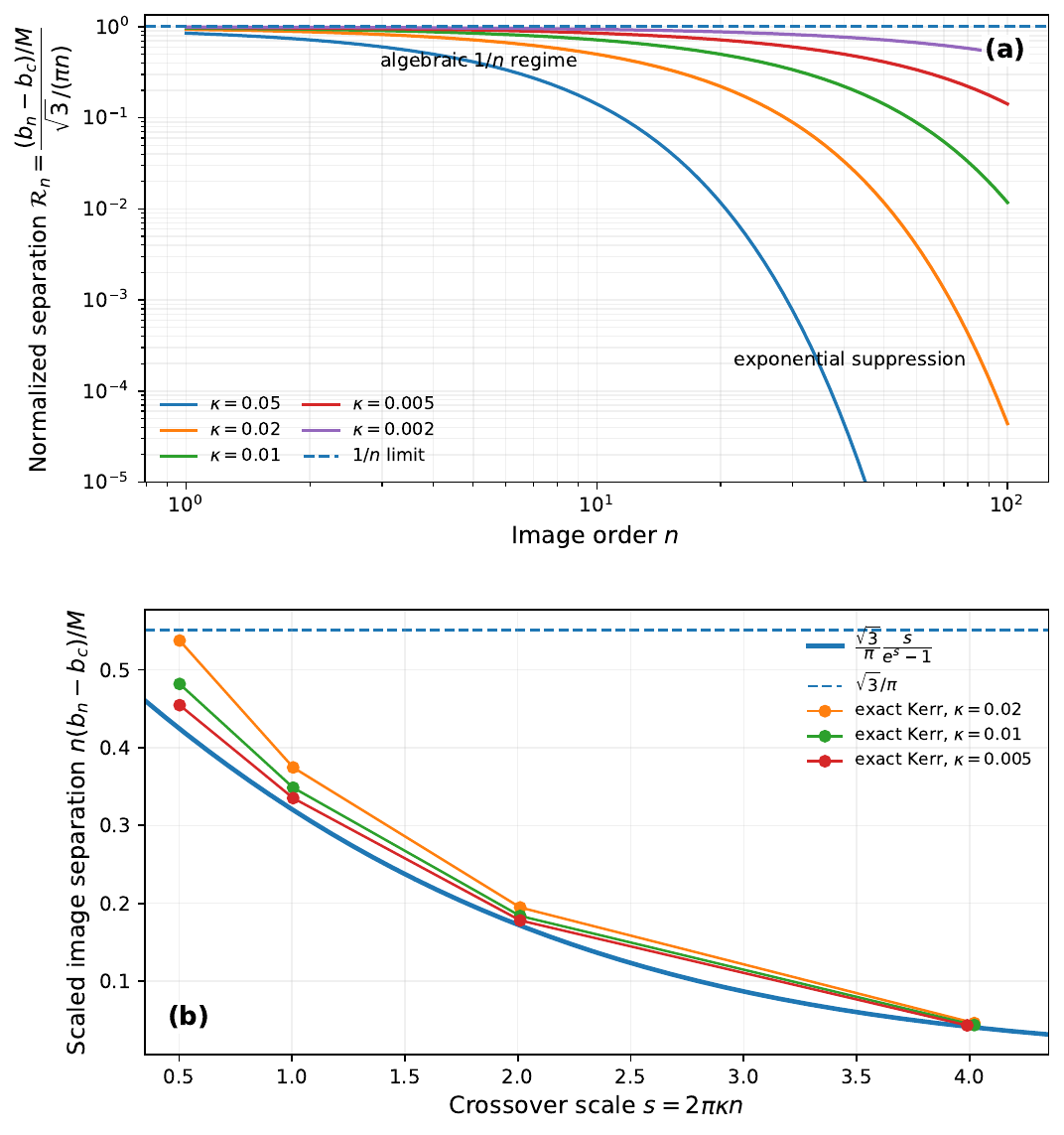}

\caption{ Validation of the higher-order image crossover for the reference choice \(\psi=0\). Panel (a) shows the image impact-parameter offset \((b_n-b_c)/M\), normalized by the extremal-like inverse-order law \(\sqrt{3}/(\pi n)\), for several values of \(\kappa\). Panel (b) compares independently solved exact Kerr image roots with the analytic master curve \((\sqrt{3}/\pi)s/[\exp(s)-1]\), using the actual scaled coordinate \(s=2\pi\kappa n\) for each integer image order. The sampled sequences were selected to lie near the target crossover scales \(s=0.5,1,2,\) and \(4\). The analytic crossover shown here illustrates the full functional form; its strong-deflection interpretation requires \(\Lcal_n\gg1\). }
\label{fig:image_crossover}
\end{figure*}

Figure~\ref{fig:image_crossover} shows the resulting crossover.
Panel (a) displays the image impact-parameter offset normalized by the
extremal-like inverse-order law, while panel (b) compares the exact
Kerr roots with the analytic master curve using the actual value of
\(s=2\pi\kappa n\) for every point.

\subsection{Exact coordinate-time delays}

The uniform delay law was tested against coordinate-time differences
computed independently from the exact Kerr geodesic equations. We used
the same discrete image branches as in the higher-order image
validation, with the reference phase choice \(\psi=0\). For each
trajectory, the same large-radius subtraction convention was applied
before image-to-image differences were formed.

The regularized coordinate times were evaluated at the three finite
outer cutoffs
\(R_{\rm max}/M=10^3,\,10^4,\,10^5\). The remaining cutoff dependence
was removed by Richardson extrapolation assuming a leading
\(R_{\rm max}^{-1}\) correction. Two extrapolations, based on the
successive cutoff pairs
\((10^3,10^4)\) and \((10^4,10^5)\), were compared as an independent
numerical convergence check, and the higher-cutoff extrapolation was
used for the final exact delay estimate.

We performed nine exact image-to-image delay comparisons using
\(\kappa=0.02,\,0.01,\,0.005\). For each value of \(\kappa\), the
integer image orders were paired so that their scaled coordinates
\(s=2\pi\kappa n\) lie near the three target crossover intervals
\(0.5\rightarrow1\), \(1\rightarrow2\), and
\(2\rightarrow4\). The actual values of
\(s_m=2\pi\kappa m\) and \(s_n=2\pi\kappa n\) were used in the
underlying exact image construction; the quoted intervals serve only
as labels for the approximately fixed-\(s\) sequences.

The exact delays were compared with the uniform prediction of
Eq.~\eqref{eq:DeltaT_nm}. We quantify the discrepancy using the
relative delay error

\begin{equation}
\epsilon_T
=
\frac{
\left|
\Delta T_{\rm exact}
-
\Delta T_{\rm uniform}
\right|
}{
\left|
\Delta T_{\rm exact}
\right|
}.
\label{eq:delay_relative_error}
\end{equation}

For each of the three target crossover intervals,
\(\epsilon_T\) decreases systematically as \(\kappa\to0\).
The residual exhibits an approximate power-law convergence,

\begin{equation}
\epsilon_T
\propto
\kappa^p,
\qquad
p
\simeq
1.91,
\label{eq:delay_order_numeric}
\end{equation}

over the range tested. The measured order is consistent with a
dominant residual close to \(\Order(\kappa^2)\), without implying an
exact quadratic asymptotic law beyond the numerical range considered.
The two independent Richardson extrapolations agree to much higher
accuracy than the finite-\(\kappa\) asymptotic errors, confirming that
the observed convergence is not controlled by the outer-cutoff
extrapolation.

\begin{figure*}[t]
\centering
\includegraphics[
width=0.82\textwidth
]{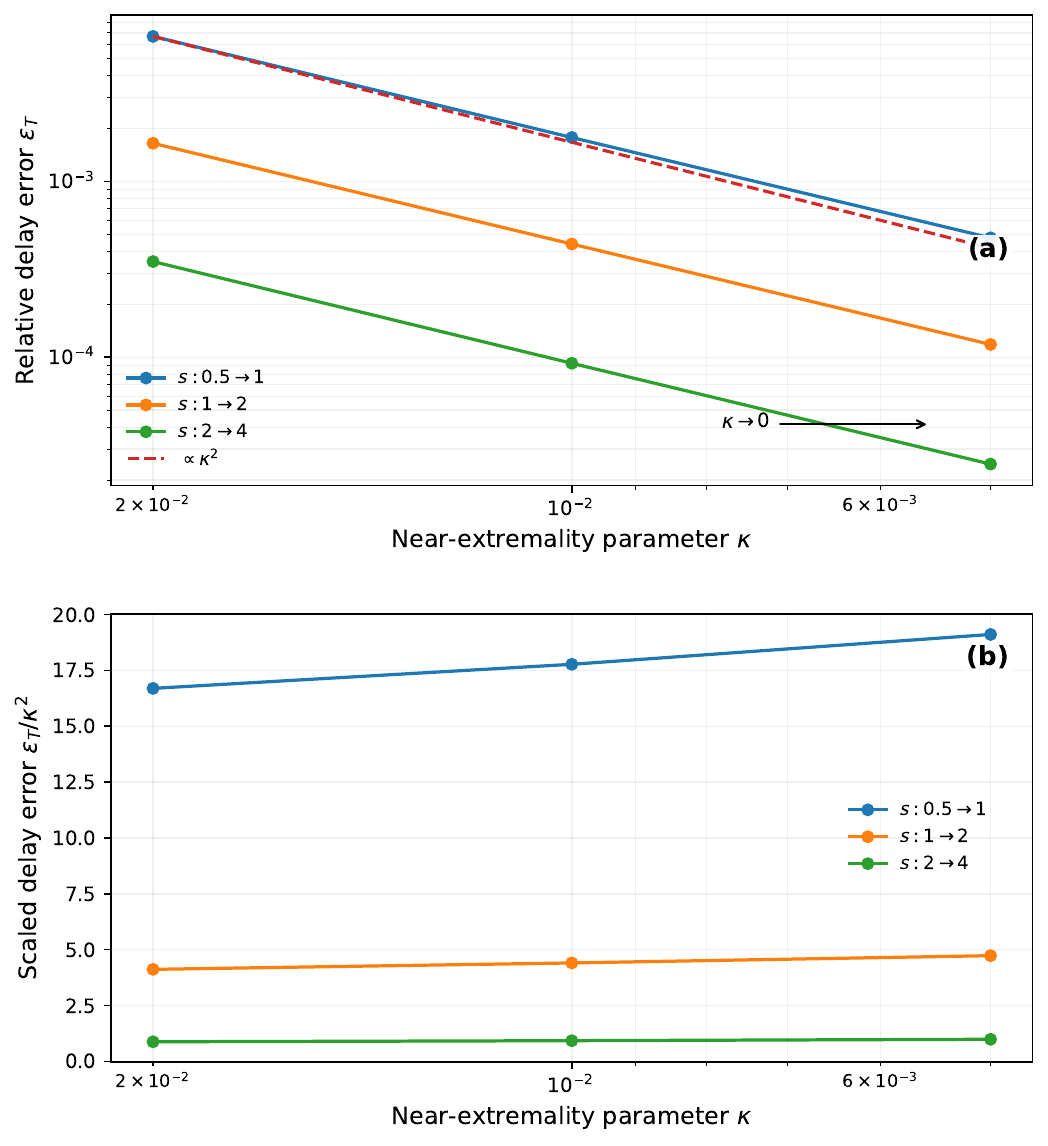}

\caption{
Validation of the uniform higher-order image time-delay law for the
reference choice \(\psi=0\).
Panel (a) shows the relative delay error
\(\epsilon_T\) of Eq.~\eqref{eq:delay_relative_error} for three
approximately fixed crossover intervals labeled by the target scales
\(s:0.5\rightarrow1\), \(1\rightarrow2\), and
\(2\rightarrow4\). The dashed line indicates a reference
\(\kappa^2\) scaling.
Panel (b) shows \(\epsilon_T/\kappa^2\); its weak remaining
dependence on \(\kappa\) is consistent with the measured convergence
order \(p\simeq1.91\).
}
\label{fig:time_delay}
\end{figure*}

\subsection{Validation summary}

The complementary numerical tests are summarized in
Table~\ref{tab:validation_summary}. Taken together, they probe four
distinct aspects of the asymptotic construction: the leading
fixed-\(\chi\) crossover, the finite next-to-leading-order matching
contribution, the joint high-order image crossover, and the
coordinate-time-delay law. In every case, the exact Kerr calculations
exhibit the convergence or improvement predicted by the corresponding
asymptotic description.

\begin{table}[t]

\caption{
Summary of the exact Kerr validation tests.
}

\label{tab:validation_summary}

\begin{ruledtabular}

\begin{tabular}{lcc}

Test
&
Cases
&
Outcome
\\

\colrule

Leading phase trajectories
&
55
&
Convergent as \(\kappa\to0\)
\\

NLO phase improvement
&
55
&
55/55 improved
\\

Fixed-\(\chi\) remainder sequences
&
11
&
11/11 convergent
\\

Exact higher-order image roots
&
12
&
12/12 improved
\\

Target-\(s\) image sequences
&
4
&
4/4 convergent
\\

Exact image-delay comparisons
&
9
&
Convergent as \(\kappa\to0\)
\\

Target-\(s\) delay intervals
&
3
&
3/3 convergent
\\

\end{tabular}

\end{ruledtabular}

\end{table}


\section{Discussion}
\label{sec:discussion}

The prograde strong-deflection problem in near-extremal Kerr is a
singular perturbation problem because the near-critical and
near-extremal limits interact. At fixed subextremal spin, the
unstable photon orbit produces the familiar logarithmic
strong-deflection divergence
\cite{Bozza2003,BozzaEtAl2005,BozzaEtAl2006,
IyerHansen2009,HsiehLeeLin2021Strong},
whereas the exactly extremal prograde problem develops a stronger
power-law critical contribution
\cite{FeleppaBozzaDeGraaf2026}. The present analysis shows that these
two limiting procedures cannot be treated independently. Their
interaction is resolved by holding the physical coordinate
\(\chi=(b-b_c)/(M\kappa)\) fixed as extremality is approached.

The origin of this scaling is geometric. The prograde critical photon
orbit approaches the outer event horizon on the radial scale
\(M\kappa\), so impact-parameter offsets of the same order probe a
region in which near-horizon and near-critical dynamics compete
simultaneously. The quadratic normal form derived in
Sec.~\ref{sec:normalform} isolates this local structure and explains
why the distinguished scale is
\(b-b_c=\Order(M\kappa)\). The uniform phase law of
Eq.~\eqref{eq:Phi_LO} then provides the asymptotic bridge between the
subextremal and extremal critical behaviors.

The higher-order image hierarchy makes the noncommutativity of the
limits particularly transparent. In the extremal-like
strong-deflection regime, successive image branches approach the
critical impact parameter algebraically, while increasing
\(\kappa\Lcal_n\) drives the sequence toward the exponential
hierarchy characteristic of ordinary subextremal lensing and
photon-ring substructure
\cite{BozzaEtAl2005,GrallaHolzWald2019,
GrallaLupsasca2020Lensing,GrallaLupsascaMarrone2020,
HadarEtAl2021,HadarKapecLupsascaStrominger2022,
BerensEtAl2026}. The transition occurs at
\(\kappa\Lcal_n=\Order(1)\), corresponding at large image order to
\(n_{\rm cross}\sim(2\pi\kappa)^{-1}\). Hence the number of image
orders that can display the algebraic hierarchy grows without bound
as \(\kappa\to0\). The double-scaling variable
\(s=2\pi\kappa n\) provides the natural description of this
competition, with Eq.~\eqref{eq:image_master} interpolating between
the two accumulation laws.

The image crossover may also have implications for horizon-scale
observations of rapidly rotating black holes. Higher-order photon-ring
subimages are generated by null geodesics that remain close to the
critical photon region, and their hierarchy can produce characteristic
signatures in long-baseline interferometric observables
\cite{JohnsonEtAl2020,GrallaLupsascaMarrone2020,
TiedeEtAl2022,CardenasAvendanoLupsasca2023,JiaEtAl2024}.
Recent studies have emphasized that photon-ring quantities such as
subimage demagnification, rotation, time delay, and interferometric
ring structure can encode information about the underlying
strong-field geometry and black-hole spin
\cite{SalehiEtAl2025,WaliaEtAl2025,
FarahEtAl2026,YanLiYue2026}.
The near-extremal crossover derived here identifies a qualitatively
different organization of the prograde equatorial near-critical
geodesics: before the ordinary exponential hierarchy is recovered,
the image branches exhibit an extremal-like algebraic accumulation
over a range whose characteristic order grows as
\(\kappa^{-1}\). This suggests that sufficiently near-extremal
rotation could leave a corresponding imprint on the high-order
photon-ring hierarchy. Establishing that connection quantitatively,
however, requires extending the present equatorial construction to
generic spherical photon orbits and mapping the resulting geodesics
onto the observer screen, including magnification and source-emission
effects.

The coordinate-time result provides an independent manifestation of
the same ordering of limits and is potentially relevant to the timing
structure of higher-order photon-ring images. Time delays between near-critical photon paths have been proposed as
complementary strong-field observables, including through photon-ring
variability, light echoes, and correlation measurements
\cite{HadarEtAl2021,WongEtAl2024,
SalehiEtAl2025,WaliaEtAl2025}.
After removal of the universal large-distance divergence,
image-to-image delays in the present equatorial problem are independent
of the additive regularization constant. In the joint high-order
extremal-like regime, neighboring branches approach
\(\Delta T_{n+1,n}\to4\pi M\). This spacing has a direct near-horizon
interpretation: since
\(\Omega_H\to(2M)^{-1}\), one additional azimuthal winding corresponds
asymptotically to a coordinate-time increment
\(2\pi/\Omega_H\to4\pi M\). The result applies when
\(n\to\infty\) while \(\kappa n\to0\), and therefore differs from
taking the high-order limit at fixed nonzero \(\kappa\).

The present construction is deliberately restricted to prograde
equatorial null geodesics. Extending it to generic Kerr lensing would
require a nonzero Carter constant and a uniform treatment of spherical
photon motion
\cite{Carter1968,Wilkins1972,Teo2003,
GrallaLupsasca2020Null}. The radial and polar motions would then be
linked through the common constants of motion, and the relevant
critical set would become a family of spherical photon orbits rather
than a single equatorial radius. Such an extension could map the
near-extremal scaling structure onto the full observer screen and
connect it directly to critical curves and photon-ring substructure
\cite{GrallaLupsasca2020Lensing,
GrallaLupsascaMarrone2020,HadarEtAl2021,
YanLiYue2026}.

A second limitation concerns the next-to-leading-order expansion.
Equation~\eqref{eq:Phi_NLO} is uniform for fixed \(\chi>0\) as
\(\kappa\to0\), but it does not reproduce the complete
fixed-subextremal next-to-leading-order coefficients arbitrarily deep
in the nested limit \(\chi\to0\). Resolving that endpoint requires an
additional inner boundary layer matched to the conventional
fixed-subextremal strong-deflection expansion. This affects the
detailed subleading coefficients but not the leading uniform law or
the fixed-\(\chi\) matched correction derived here.

Finally, the image condition used in this work isolates the local
near-critical phase structure rather than constructing a complete
finite-distance Kerr lens map. Observable lensing predictions require
additional information about the source and observer positions,
observer-screen coordinates, magnifications, and the global
source--image mapping
\cite{BozzaScarpetta2007,GrallaLupsasca2020Lensing,
GrallaLupsascaMarrone2020,
RojasPaterninaCardenasAvendano2026}.
The local strong-deflection structure derived here can serve as an
analytic building block for such a construction, while complete
observer-screen predictions require those additional global
ingredients.


\section{Conclusions}
\label{sec:conclusions}

We have developed a uniform strong-deflection expansion for prograde
equatorial Kerr null geodesics in the joint near-extremal and
near-critical limit. The distinguished scaling is governed by the
physical crossover coordinate
\(\chi=(b-b_c)/(M\kappa)\), for which the impact-parameter offset from
criticality is of the same order as the near-extremal scale
\(M\kappa\).

In this limit, the Kerr radial potential reduces at leading order to a
quadratic normal form, and the leading azimuthal accumulation is

\begin{equation}
\Phi_{\rm LO}
=
\frac{1}{\kappa}
\ln
\left(
1+\frac{2\sqrt{3}}{\chi}
\right).
\label{eq:conclusion_uniform}
\end{equation}

This single expression connects the near-extremal form of the
logarithmic strong-deflection behavior of subextremal Kerr to the
power-law critical behavior of the exactly extremal prograde problem.
Thus the logarithmic and power-law behaviors are unified as two
asymptotic limits of the same near-extremal scaling law.

We extended the expansion to next-to-leading order by matching the
near-horizon and outer Kerr regions and recentering the result on the
exact critical impact parameter. Direct comparison with high-precision
exact Kerr geodesics shows that the NLO approximation improves on the
leading result for all 55 trajectories tested, with improvement factors
ranging from approximately \(6.35\) to \(9.81\times10^2\). An
independent extraction of the finite matched remainder converges toward
the analytic constant \(C_{\phi}\), providing a separate validation of
the subleading structure.

The same uniform structure determines the higher-order image
hierarchy. We find

\begin{equation}
\frac{b_n-b_c}{M}
=
\frac{
2\sqrt{3}\,\kappa
}{
\exp(\kappa\Lcal_n)-1
},
\label{eq:conclusion_image}
\end{equation}

which connects algebraic inverse-order accumulation on the
extremal-like side of the crossover with the exponential dependence
characteristic of ordinary subextremal Kerr. The crossover is
controlled by \(\kappa\Lcal_n\), or equivalently by \(\kappa n\) at
large image order, and occurs at the characteristic scale

\begin{equation}
n_{\rm cross}
\sim
\frac{1}{2\pi\kappa}.
\end{equation}

Thus the number of image orders that can exhibit extremal-like
algebraic accumulation grows parametrically as extremality is
approached. The corresponding double-scaling limit,
\(\kappa\to0\) and \(n\to\infty\) with
\(s=2\pi\kappa n=\Order(1)\), is described by the parameter-free
master curve derived in Sec.~\ref{sec:images}.

We also derived the corresponding uniform image-to-image
coordinate-time-delay law. Independent exact Kerr time integrations
converge systematically toward the analytic prediction, with a
measured residual scaling \(p\simeq1.91\), consistent over the tested
range with a dominant correction close to \(\Order(\kappa^2)\).
In the joint high-order extremal-like regime, neighboring image
branches approach the coordinate-time spacing
\(\Delta T_{n+1,n}\to4\pi M\).

Taken together, these results provide a unified asymptotic description
of the singular transition between subextremal and exactly extremal
prograde Kerr strong deflection. The present construction is restricted
to equatorial motion and isolates the local near-critical phase
structure. A natural next step is to extend the uniform near-extremal
analysis to generic spherical photon orbits and map the resulting
geodesic hierarchy onto the observer screen. Such an extension would
allow the crossover derived here to be connected quantitatively with
photon-ring separations, demagnification, rotation, and time-delay
observables relevant to future horizon-scale interferometric studies.


\section*{Data Availability}

The data and numerical results supporting the findings of this study
are available from the author upon reasonable request.

\appendix


\section{Equatorial Kerr Geodesics, Critical Orbit, and Normal Form}
\label{app:geodesics}

This appendix collects the exact equatorial geodesic relations and
near-extremal expansions used in the main text. We use the conventions
and definitions introduced in Secs.~\ref{sec:geodesics} and
\ref{sec:normalform}.

\subsection{Equatorial first-order equations}

Let \(p_\mu\) denote the covariant photon momentum. Stationarity and
axial symmetry imply the conserved quantities

\begin{equation}
E
=
-p_t,
\qquad
L_z
=
p_\phi.
\end{equation}

The separability of the Kerr geodesic equations yields the additional
Carter constant
\cite{Carter1968,Wilkins1972,GrallaLupsasca2020Null}.
For equatorial motion, \(\theta=\pi/2\), the Carter constant vanishes.

Defining the energy-normalized azimuthal impact parameter and radial
combination by

\begin{equation}
b
=
\frac{L_z}{E},
\qquad
P(r)
=
r^2+a^2-ab,
\end{equation}

the energy-normalized radial potential is

\begin{equation}
R(r)
=
P(r)^2
-
\Delta(b-a)^2.
\label{eq:R_app_exact}
\end{equation}

Since \(\Sigma=r^2\) in the equatorial plane, the separated
first-order equations become

\begin{align}
\dot r
&=
\pm
\frac{E}{r^2}\sqrt{R},
\label{eq:rdot_app}
\\
\dot\phi
&=
\frac{E}{r^2}
\left[
(b-a)
+
\frac{aP}{\Delta}
\right],
\label{eq:phidot_app}
\\
\dot t
&=
\frac{E}{r^2}
\left[
a(b-a)
+
\frac{(r^2+a^2)P}{\Delta}
\right].
\label{eq:tdot_app}
\end{align}

On the outgoing radial branch, \(\dot r>0\). Dividing the azimuthal
and temporal equations by Eq.~\eqref{eq:rdot_app} gives

\begin{align}
\frac{\dd\phi}{\dd r}
&=
\frac{
(b-a)+aP/\Delta
}{
\sqrt{R}
},
\label{eq:dphi_dr_app}
\\
\frac{\dd t}{\dd r}
&=
\frac{
a(b-a)+(r^2+a^2)P/\Delta
}{
\sqrt{R}
}.
\label{eq:dt_dr_app}
\end{align}

These exact expressions are used in both the asymptotic analysis and
the independent numerical calculations.

\subsection{Exact turning-point relation}

Let \(r_0\) denote the outer radial turning point of a scattering
trajectory. It satisfies

\begin{equation}
R(r_0)
=
0.
\end{equation}

Writing

\begin{equation}
B
=
b-a,
\end{equation}

the radial combination becomes

\begin{equation}
P
=
r^2-aB.
\end{equation}

On the physical prograde scattering branch, the turning-point
condition may be written as

\begin{equation}
P_0
=
\sqrt{\Delta_0}\,B,
\end{equation}

where

\begin{equation}
\Delta_0
\equiv
\Delta(r_0).
\end{equation}

Thus

\begin{equation}
r_0^2-aB
=
\sqrt{\Delta_0}\,B,
\end{equation}

which gives the exact relation

\begin{equation}
b
=
a
+
\frac{
r_0^2
}{
a+\sqrt{\Delta_0}
},
\qquad
\Delta_0
=
r_0^2-2Mr_0+a^2.
\label{eq:turning_app_exact}
\end{equation}

Equation~\eqref{eq:turning_app_exact} is the starting point for the
next-to-leading-order turning-point expansion developed in
Appendix~\ref{app:nlo}.

\subsection{Critical photon orbit}
\label{app:critical}

At the unstable equatorial circular photon orbit, the radial turning
point becomes a double root,

\begin{equation}
R(r_c)
=
0,
\qquad
R_{,r}(r_c)
=
0,
\label{eq:critical_app_conditions}
\end{equation}

where \(R_{,r}\equiv\partial R/\partial r\) is evaluated at fixed
impact parameter \(b\).

Since

\begin{equation}
R_{,r}
=
4rP
-
2(r-M)(b-a)^2,
\end{equation}

combining the double-root condition with

\begin{equation}
P_c
=
\sqrt{\Delta_c}\,(b_c-a)
\end{equation}

eliminates the impact parameter and gives the standard prograde
equatorial photon-orbit equation,

\begin{equation}
r_c^2
-
3Mr_c
+
2a\sqrt{Mr_c}
=
0.
\label{eq:photon_orbit_app}
\end{equation}

Its prograde solution is

\begin{equation}
\frac{r_c}{M}
=
2
\left[
1+
\cos
\left(
\frac{2}{3}
\cos^{-1}
\left(
-\frac{a}{M}
\right)
\right)
\right].
\label{eq:rc_app_exact}
\end{equation}

Substitution back into the turning-point relation gives the exact
critical impact parameter,

\begin{equation}
b_c
=
a
+
\frac{
r_c^{3/2}
}{
\sqrt{M}
}.
\label{eq:bc_app_exact}
\end{equation}

\subsection{Near-extremal critical-orbit expansion}

Introduce an auxiliary angle \(\vartheta\) by

\begin{equation}
\frac{a}{M}
=
\cos\vartheta,
\qquad
\kappa
=
\sin\vartheta.
\end{equation}

For the near-extremal prograde branch,
\(\vartheta\ll1\), and therefore

\begin{equation}
\cos^{-1}
\left(
-\frac{a}{M}
\right)
=
\pi-\vartheta.
\end{equation}

Equation~\eqref{eq:rc_app_exact} then becomes

\begin{equation}
\frac{r_c}{M}
=
2
\left[
1+
\cos
\left(
\frac{2}{3}
(\pi-\vartheta)
\right)
\right].
\end{equation}

Using

\begin{equation}
\vartheta
=
\arcsin\kappa
=
\kappa
+
\frac{\kappa^3}{6}
+
\Order(\kappa^5),
\end{equation}

one obtains

\begin{align}
\frac{r_c}{M}
&=
1
+
\frac{2}{\sqrt{3}}\kappa
+
\frac{2}{9}\kappa^2
+
\frac{5}{27\sqrt{3}}\kappa^3
+
\Order(\kappa^4),
\nonumber\\
\frac{b_c}{M}
&=
2
+
\sqrt{3}\,\kappa
+
\frac{1}{3}\kappa^2
+
\frac{4\sqrt{3}}{27}\kappa^3
+
\Order(\kappa^4).
\label{eq:critical_series_app}
\end{align}

Since the outer horizon satisfies

\begin{equation}
\frac{r_+}{M}
=
1+\kappa,
\end{equation}

the critical orbit remains outside the event horizon by

\begin{equation}
\frac{r_c-r_+}{M}
=
\left(
\frac{2}{\sqrt{3}}-1
\right)\kappa
+
\Order(\kappa^2).
\label{eq:critical_horizon_app}
\end{equation}

Thus the prograde critical orbit enters the near-horizon region on the
same radial scale \(M\kappa\) that controls the horizon separation.

\subsection{Near-horizon radial normal form}
\label{app:normal_form}

We introduce the scaled variables

\begin{equation}
\frac{r}{M}
=
1+\kappa y,
\qquad
\frac{b}{M}
=
2+\kappa X.
\label{eq:scaled_app}
\end{equation}

Here \(y\) resolves the near-horizon radial coordinate, while \(X\) is
the intermediate leading-order scaled impact-parameter coordinate
introduced in Sec.~\ref{sec:normalform}.

Using

\begin{equation}
\frac{a}{M}
=
\sqrt{1-\kappa^2},
\end{equation}

the horizon function becomes exactly

\begin{equation}
\frac{\Delta}{M^2}
=
\kappa^2(y^2-1).
\label{eq:Delta_app_scaled}
\end{equation}

The radial combination \(P\) expands as

\begin{equation}
\frac{P}{M^2}
=
\kappa(2y-X)
+
\kappa^2y^2
+
\Order(\kappa^3),
\label{eq:P_app_scaled}
\end{equation}

while

\begin{equation}
\frac{b-a}{M}
=
1
+
\kappa X
+
\Order(\kappa^2).
\end{equation}

Substitution into Eq.~\eqref{eq:R_app_exact} gives

\begin{equation}
\frac{R}{M^4}
=
\kappa^2
Q(y;X)
+
\Order(\kappa^3),
\label{eq:R_app_normal}
\end{equation}

where

\begin{equation}
Q(y;X)
=
3y^2-4Xy+X^2+1.
\label{eq:Q_app}
\end{equation}

The two reduced radial roots are

\begin{equation}
y_\pm
=
\frac{
2X\pm D
}{
3
},
\qquad
D
\equiv
\sqrt{X^2-3}.
\label{eq:roots_app}
\end{equation}

For the scattering branch considered here,
\(X>\sqrt{3}\) and hence \(D>0\). The roots merge at the reduced
critical point,

\begin{equation}
X_c
=
\sqrt{3},
\qquad
y_c
=
\frac{2}{\sqrt{3}},
\end{equation}

in agreement with the leading terms of
Eq.~\eqref{eq:critical_series_app}.

\subsection{Distinguished critical scaling}

The reduced root separation is

\begin{equation}
y_+-y_-
=
\frac{2}{3}
\sqrt{X^2-3}.
\end{equation}

Writing

\begin{equation}
X
=
\sqrt{3}+\delta X,
\end{equation}

gives

\begin{equation}
y_+-y_-
=
\frac{2}{3}
\sqrt{
2\sqrt{3}\,\delta X
+
\Order(\delta X^2)
}.
\end{equation}

The exact critical expansion in
Eq.~\eqref{eq:critical_series_app}, together with
\(b/M=2+\kappa X\), gives

\begin{equation}
\frac{
b-b_c
}{
M\kappa
}
=
X-\sqrt{3}
+
\Order(\kappa).
\label{eq:chi_X_app}
\end{equation}

Thus the nontrivial joint near-critical and near-extremal limit is
obtained by holding the physical crossover coordinate

\begin{equation}
\chi
\equiv
\frac{
b-b_c
}{
M\kappa
}
\end{equation}

fixed as \(\kappa\to0\). Consequently,

\begin{equation}
X
=
\sqrt{3}
+
\chi
+
\Order(\kappa).
\label{eq:X_chi_app}
\end{equation}

At leading order this reduces to
\(X=\sqrt{3}+\chi\), while the \(\Order(\kappa)\) correction becomes
important when the expansion is recentered on the exact critical
impact parameter at next-to-leading order in Appendix~\ref{app:nlo}.


\section{Evaluation of the Leading Crossover Integral}
\label{app:F}

This appendix gives the intermediate steps leading from the exact Kerr
azimuthal integral to the closed leading crossover function.

\subsection{Leading azimuthal integrand}

Using Eqs.~\eqref{eq:dphi_dr_app} and
\eqref{eq:scaled_app}, together with
\(\dd r=M\kappa\,\dd y\), the azimuthal differential becomes

\begin{equation}
\dd\phi
=
\frac{1}{\kappa}
\frac{
2y-X
}{
(y^2-1)\sqrt{Q(y;X)}
}
\,\dd y
+
\Order(1).
\label{eq:dphi_app_scaled}
\end{equation}

The leading two-leg contribution may therefore be written as

\begin{equation}
\Phi_{\rm LO}
=
\frac{1}{\kappa}
F_X(X),
\label{eq:Phi_FX_app}
\end{equation}

where the reduced crossover function is

\begin{equation}
F_X(X)
=
2
\int_{y_+}^{\infty}
\frac{
2y-X
}{
(y^2-1)\sqrt{Q(y;X)}
}
\,\dd y.
\label{eq:F_integral_app}
\end{equation}

For the physical scattering branch,
\(X>\sqrt{3}\), so the outer reduced turning point
\(y_+\) lies outside the double-root critical point.

\subsection{Reduction to elementary integrals}

We first decompose

\begin{equation}
\frac{2y-X}{y^2-1}
=
\frac{2-X}{2(y-1)}
+
\frac{2+X}{2(y+1)}.
\label{eq:partial_fraction_app}
\end{equation}

For \(c=\pm1\), define

\begin{equation}
I_c
=
\int_{y_+}^{\infty}
\frac{
\dd y
}{
(y-c)\sqrt{Q(y;X)}
}.
\end{equation}

Using the substitution

\begin{equation}
u^2
=
\frac{
y-y_+
}{
y-y_-
},
\label{eq:u_sub_app}
\end{equation}

one obtains

\begin{equation}
y
=
\frac{
y_+-u^2y_-
}{
1-u^2
},
\end{equation}

together with

\begin{equation}
\frac{\dd y}{\sqrt{Q}}
=
\frac{
2\,\dd u
}{
\sqrt{3}(1-u^2)
}.
\end{equation}

Moreover,

\begin{equation}
y-c
=
\frac{
(y_+-c)-u^2(y_--c)
}{
1-u^2
}.
\end{equation}

Hence

\begin{equation}
I_c
=
\frac{2}{\sqrt{3}}
\int_0^1
\frac{
\dd u
}{
(y_+-c)-u^2(y_--c)
}.
\label{eq:Ic_app}
\end{equation}

The remaining integral is elementary. Depending on the sign of
\((y_+-c)(y_--c)\), it may be represented by a real
\(\operatorname{artanh}\) or \(\arctan\) expression. Combining the
\(c=1\) and \(c=-1\) contributions before simplification avoids the
need to treat the two representations separately.

Using

\begin{equation}
Q(1)
=
(X-2)^2,
\qquad
Q(-1)
=
(X+2)^2,
\end{equation}

the two contributions generated by
Eq.~\eqref{eq:partial_fraction_app} combine to give

\begin{equation}
F_X(X)
=
2
\operatorname{artanh}
\left(
\frac{\sqrt{3}}{X}
\right),
\qquad
X>\sqrt{3}.
\end{equation}

Using

\begin{equation}
2\operatorname{artanh}z
=
\ln
\left(
\frac{1+z}{1-z}
\right),
\end{equation}

we obtain

\begin{equation}
F_X(X)
=
\ln
\left(
\frac{
X+\sqrt{3}
}{
X-\sqrt{3}
}
\right),
\qquad
X>\sqrt{3}.
\label{eq:F_result_app}
\end{equation}

Using the leading relation

\begin{equation}
X
=
\sqrt{3}+\chi,
\end{equation}

we define the corresponding physical crossover function by

\begin{equation}
F(\chi)
\equiv
F_X(\sqrt{3}+\chi)
=
\ln
\left(
1+\frac{
2\sqrt{3}
}{
\chi
}
\right),
\qquad
\chi>0.
\label{eq:F_chi_app}
\end{equation}

The leading uniform phase is therefore

\begin{equation}
\Phi_{\rm LO}
=
\frac{
F(\chi)
}{
\kappa
}.
\label{eq:Phi_LO_app}
\end{equation}

\subsection{Endpoint limits}

On the near-critical side of the crossover,
\(\chi\ll1\),

\begin{equation}
F(\chi)
=
-\ln\chi
+
\ln(2\sqrt{3})
+
\Order(\chi).
\label{eq:F_small_app}
\end{equation}

Hence

\begin{equation}
\Phi_{\rm LO}
=
\frac{1}{\kappa}
\left[
-\ln\chi
+
\ln(2\sqrt{3})
+
\Order(\chi)
\right],
\end{equation}

which gives the near-extremal form of the logarithmic
strong-deflection singularity.

On the opposite side, in the extremal-like overlap

\begin{equation}
1
\ll
\chi
\ll
\kappa^{-1},
\label{eq:chi_large_app}
\end{equation}

the crossover function satisfies

\begin{equation}
F(\chi)
=
\frac{
2\sqrt{3}
}{
\chi
}
+
\Order(\chi^{-2}),
\label{eq:F_large_app}
\end{equation}

and therefore

\begin{equation}
\Phi_{\rm LO}
\sim
\frac{
2\sqrt{3}
}{
\kappa\chi
}
=
\frac{
2\sqrt{3}M
}{
b-b_c
}.
\label{eq:Phi_power_app}
\end{equation}

These endpoint expansions reproduce the logarithmic and power-law
limits discussed in Sec.~\ref{sec:leading}.


\section{Derivation of the Matched Next-to-Leading-Order Expansion}\label{app:nlo}

This appendix gives the intermediate steps leading to
Eq.~\eqref{eq:Phi_NLO}, including the turning-point expansion,
the subleading radial potential and azimuthal integrand, recentering
on the exact critical impact parameter, the near--outer matching,
and the cancellation of the intermediate matching scale.

\subsection{Turning-point expansion}

Write the outer turning point as

\begin{equation}
\frac{r_0}{M}
=
1+\kappa y_0,
\qquad
s_0
\equiv
\sqrt{y_0^2-1}.
\label{eq:r0_s0_nlo_app}
\end{equation}

At the turning point,

\begin{equation}
\frac{\sqrt{\Delta_0}}{M}
=
\kappa s_0,
\end{equation}

while the near-extremal spin expansion is

\begin{equation}
\frac{a}{M}
=
1-\frac{\kappa^2}{2}
+
\Order(\kappa^4).
\end{equation}

Using the exact turning-point relation
Eq.~\eqref{eq:turning_app_exact},

\begin{equation}
\frac{b}{M}
=
\frac{a}{M}
+
\frac{
(1+\kappa y_0)^2
}{
a/M+\kappa s_0
}.
\end{equation}

The reciprocal denominator expands as

\begin{equation}
\frac{1}{
a/M+\kappa s_0
}
=
1
-
\kappa s_0
+
\kappa^2
\left(
s_0^2+\frac{1}{2}
\right)
+
\Order(\kappa^3).
\end{equation}

Combining these expressions gives

\begin{equation}
\frac{b}{M}
=
2
+
\kappa X
+
\kappa^2\beta
+
\Order(\kappa^3),
\label{eq:b_NLO_app}
\end{equation}

where

\begin{equation}
X
=
2y_0-s_0,
\qquad
\beta
=
2y_0^2-2y_0s_0-1.
\label{eq:Xbeta_app}
\end{equation}

Since \(s_0=2y_0-X\), squaring this relation gives

\begin{equation}
3y_0^2
-
4Xy_0
+
X^2
+
1
=
0.
\end{equation}

The physical outer root is therefore

\begin{equation}
y_0
=
\frac{
2X+D
}{
3
},
\qquad
D
\equiv
\sqrt{X^2-3}.
\label{eq:y0_NLO_app}
\end{equation}

Substitution into Eq.~\eqref{eq:Xbeta_app} gives

\begin{equation}
\beta(X)
=
\frac{
2X^2-2XD-3
}{
9
}.
\label{eq:beta_app}
\end{equation}

At the reduced critical point,
\(X=\sqrt{3}\) and \(D=0\), so

\begin{equation}
\beta_c
\equiv
\beta(\sqrt{3})
=
\frac{1}{3},
\end{equation}

in agreement with the exact near-extremal critical expansion in
Eq.~\eqref{eq:critical_series_app}.

\subsection{Subleading radial potential}

With

\begin{equation}
\frac{b}{M}
=
2+\kappa X+\kappa^2\beta+\cdots,
\end{equation}

the radial combination \(P\) becomes

\begin{equation}
\frac{P}{M^2}
=
\kappa(2y-X)
+
\kappa^2(y^2-\beta)
+
\Order(\kappa^3).
\label{eq:P_NLO_app}
\end{equation}

Similarly,

\begin{equation}
\frac{b-a}{M}
=
1+\kappa X
+
\kappa^2
\left(
\beta+\frac{1}{2}
\right)
+
\Order(\kappa^3).
\end{equation}

Substituting these expansions into the exact radial potential gives

\begin{equation}
\frac{R}{M^4}
=
\kappa^2Q
+
\kappa^3R_3
+
\Order(\kappa^4),
\label{eq:R_NLO_app}
\end{equation}

where

\begin{equation}
R_3
=
2(2y-X)(y^2-\beta)
-
2X(y^2-1),
\label{eq:R3_app}
\end{equation}

and

\begin{equation}
Q(y;X)
=
3y^2-4Xy+X^2+1
\end{equation}

is the leading reduced potential introduced in
Appendix~\ref{app:normal_form}.

\subsection{Subleading azimuthal integrand}

Expanding Eq.~\eqref{eq:dphi_dr_app} consistently through the next
order gives

\begin{equation}
\frac{\dd\phi}{\dd y}
=
\frac{1}{\kappa}
f_0(y;X)
+
f_1(y;X)
+
\Order(\kappa),
\label{eq:phi_integrand_NLO_app}
\end{equation}

with

\begin{align}
f_0
&=
\frac{
2y-X
}{
(y^2-1)\sqrt{Q}
},
\nonumber\\
f_1
&=
\frac{
2y^2-2Xy+\beta+1
}{
Q^{3/2}
}.
\label{eq:f01_app}
\end{align}

The subleading term admits the decomposition

\begin{equation}
f_1
=
\frac{\dd}{\dd y}
\left(
\frac{
Ay+B
}{
\sqrt{Q}
}
\right)
+
\frac{2}{3\sqrt{Q}},
\label{eq:f1_decomp_app}
\end{equation}

where

\begin{equation}
A
=
\frac{2X}{3D},
\qquad
B
=
-
\frac{
2X(X^2+2XD-3)
}{
9D^2
}.
\label{eq:AB_app}
\end{equation}

At the physical turning point,

\begin{equation}
Ay_0+B
=
0.
\label{eq:endpoint_cancel_app}
\end{equation}

Thus the total-derivative contribution has no lower-endpoint
contribution at the physical turning point.

The remaining elementary integral follows from

\begin{equation}
Q
=
\frac{
(3y-2X)^2-D^2
}{
3
},
\end{equation}

which gives

\begin{equation}
\int_{y_0}^{Y}
\frac{\dd y}{\sqrt{Q}}
=
\frac{1}{\sqrt{3}}
\operatorname{arcosh}
\left(
\frac{
3Y-2X
}{
D
}
\right).
\label{eq:Q_integral_app}
\end{equation}

For \(Y\gg1\),

\begin{equation}
\int_{y_0}^{Y}
\frac{\dd y}{\sqrt{Q}}
=
\frac{1}{\sqrt{3}}
\ln
\left(
\frac{6Y}{D}
\right)
+
\Order(Y^{-1}).
\label{eq:Q_largeY_app}
\end{equation}

\subsection{Recentering on the physical critical impact parameter}

The variable \(X\) appearing in the intermediate near-horizon
expansion is centered on the leading critical value but not yet on
the exact Kerr critical impact parameter beyond leading order. The
physical crossover coordinate is

\begin{equation}
\chi
=
\frac{
b-b_c
}{
M\kappa
},
\end{equation}

and we define

\begin{equation}
x
\equiv
\sqrt{3}+\chi.
\label{eq:x_physical_app}
\end{equation}

Using Eq.~\eqref{eq:b_NLO_app} together with the exact critical
expansion in Eq.~\eqref{eq:critical_series_app} gives

\begin{equation}
x
=
X
+
\kappa
\left(
\beta-\frac{1}{3}
\right)
+
\Order(\kappa^2).
\label{eq:x_X_relation_app}
\end{equation}

Hence

\begin{equation}
X
=
x
-
\kappa
\left(
\beta-\frac{1}{3}
\right)
+
\Order(\kappa^2).
\label{eq:X_recenter_app}
\end{equation}

It is useful to define the discriminant associated with the physical
coordinate \(x\) by

\begin{equation}
D_x
\equiv
\sqrt{x^2-3}.
\label{eq:Dx_app}
\end{equation}

Since \(X=x+\Order(\kappa)\), one has, at fixed
\(x>\sqrt{3}\),

\begin{equation}
D
=
D_x+\Order(\kappa).
\label{eq:D_Dx_app}
\end{equation}

Consequently, replacing \(D\) by \(D_x\) in terms that already enter
at order unity changes the expansion only beyond the order retained
here.

The reduced crossover function introduced in
Eq.~\eqref{eq:F_X} satisfies

\begin{equation}
F_X'(x)
=
-
\frac{
2\sqrt{3}
}{
x^2-3
}
=
-
\frac{
2\sqrt{3}
}{
D_x^2
}.
\label{eq:Fprime_x_app}
\end{equation}

Expanding the leading contribution about the physical variable \(x\)
gives

\begin{align}
\frac{F_X(X)}{\kappa}
={}&
\frac{F_X(x)}{\kappa}
-
\left(
\beta-\frac{1}{3}
\right)
F_X'(x)
+
\Order(\kappa).
\label{eq:F_recenter_app}
\end{align}

At the order required in this finite contribution,
\(\beta\) may be evaluated at \(x\). From
Eq.~\eqref{eq:beta_app},

\begin{equation}
\beta(x)-\frac{1}{3}
=
-
\frac{
2D_x(x-D_x)
}{
9
}.
\label{eq:beta_recenter_app}
\end{equation}

The resulting recentering contribution is

\begin{equation}
-
\left(
\beta-\frac{1}{3}
\right)
F_X'(x)
=
-
\frac{
4\sqrt{3}\,x
}{
9D_x
}
+
\frac{
4\sqrt{3}
}{
9
}.
\label{eq:recenter_cancel_app}
\end{equation}

The term proportional to \(x/D_x\) cancels the corresponding
endpoint contribution generated by the total derivative in
Eq.~\eqref{eq:f1_decomp_app}. This cancellation is essential for
expressing the matched result in terms of the physical displacement
from the exact critical impact parameter.

\subsection{Near--outer matching}

Introduce an intermediate matching point

\begin{equation}
\frac{r_{\rm m}}{M}
=
1+\kappa Y,
\end{equation}

with

\begin{equation}
1
\ll
Y
\ll
\kappa^{-1}.
\label{eq:matching_window_app}
\end{equation}

Using Eqs.~\eqref{eq:f1_decomp_app}--\eqref{eq:recenter_cancel_app},
the two-leg near-region contribution has the large-\(Y\) form

\begin{align}
\Phi_{\rm near}
={}&
\frac{F_X(x)}{\kappa}
-
\frac{
4
}{
\sqrt{3}\,\kappa Y
}
\nonumber\\
&+
\frac{
4\sqrt{3}
}{
9
}
\left[
1+
\ln
\left(
\frac{
6Y
}{
D_x
}
\right)
\right]
+
o(1).
\label{eq:Phi_near_match_app}
\end{align}

For the outer contribution, the extremal values
\(a=M\) and \(b=2M\) may be used at the order required here. Setting
\(M=1\) temporarily for the integration,
Eq.~\eqref{eq:dphi_dr_app} becomes

\begin{equation}
\left.
\frac{\dd\phi}{\dd r}
\right|_{\rm ext}
=
\frac{
2r
}{
(r-1)^2
\sqrt{r(r+2)}
}.
\label{eq:outer_ext_integrand_app}
\end{equation}

The two-leg outer integral is therefore

\begin{equation}
\Phi_{\rm out}
=
4
\int_{r_{\rm m}}^{\infty}
\frac{
r\,\dd r
}{
(r-1)^2\sqrt{r(r+2)}
}.
\label{eq:outer_integral_app}
\end{equation}

Using

\begin{equation}
z
=
\sqrt{
\frac{r}{r+2}
},
\end{equation}

the integral becomes rational,

\begin{equation}
\Phi_{\rm out}
=
16
\int_{z_{\rm m}}^1
\frac{
z^2
}{
(3z^2-1)^2
}
\,\dd z.
\label{eq:outer_integral_z_app}
\end{equation}

Writing, in these temporarily dimensionless units,

\begin{equation}
\epsilon
=
r_{\rm m}-1
=
\kappa Y,
\end{equation}

its small-\(\epsilon\) expansion is

\begin{align}
\Phi_{\rm out}
={}&
\frac{
4
}{
\sqrt{3}\,\epsilon
}
-
\frac{
4\sqrt{3}
}{
9
}
\ln\epsilon
+
C_{\rm out}
+
\Order(\epsilon),
\label{eq:Phi_outer_expand_app}
\end{align}

with

\begin{equation}
C_{\rm out}
=
\frac{
8\sqrt{3}
}{
9
}
+
\frac{
4\sqrt{3}
}{
9
}
\ln
\left[
6(2-\sqrt{3})
\right]
-
\frac{4}{3}.
\label{eq:Cout_app}
\end{equation}

Substituting \(\epsilon=\kappa Y\) into
Eq.~\eqref{eq:Phi_outer_expand_app} and adding
Eq.~\eqref{eq:Phi_near_match_app}, the terms proportional to
\(1/(\kappa Y)\) cancel. All logarithmic dependence on the arbitrary
matching scale \(Y\) cancels as well.

The matched result is therefore

\begin{align}
\Phi
={}&
\frac{F_X(x)}{\kappa}
-
\frac{
4\sqrt{3}
}{
9
}
\ln(\kappa D_x)
+
C_\phi
+
o(1),
\label{eq:Phi_match_app}
\end{align}

where

\begin{align}
C_\phi
&=
C_{\rm out}
+
\frac{
4\sqrt{3}
}{
9
}
\left(
1+\ln 6
\right)
\nonumber\\
&=
\frac{
4\sqrt{3}
}{
9
}
\left[
3+
\ln
\left(
36(2-\sqrt{3})
\right)
\right]
-
\frac{4}{3}.
\label{eq:Cphi_derivation_app}
\end{align}

Numerically,

\begin{equation}
C_\phi
=
2.7208672467\ldots.
\label{eq:Cphi_numeric_app}
\end{equation}

Since

\begin{equation}
F_X(x)
=
\ln
\left(
\frac{
x+\sqrt{3}
}{
x-\sqrt{3}
}
\right),
\qquad
D_x
=
\sqrt{x^2-3},
\end{equation}

Eq.~\eqref{eq:Phi_match_app} reproduces
Eq.~\eqref{eq:Phi_NLO} in the main text. The explicit disappearance
of the intermediate scale \(Y\) provides a direct consistency check
of the near--outer matching.

\subsection{Extremal-like overlap}

To connect the uniform near-extremal expansion with the exactly
extremal Kerr strong-deflection limit, define the dimensionless
impact-parameter offset from the extremal critical value,

\begin{equation}
\delta_b
\equiv
\frac{
b-2M
}{
M
}.
\label{eq:delta_b_app}
\end{equation}

Using

\begin{equation}
b
=
b_c+M\kappa\chi
\end{equation}

together with the critical expansion
Eq.~\eqref{eq:critical_series_app}, one obtains

\begin{equation}
\delta_b
=
\kappa
\left(
\chi+\sqrt{3}
\right)
+
\frac{1}{3}\kappa^2
+
\Order(\kappa^3).
\label{eq:delta_b_chi_app}
\end{equation}

The extremal-like overlap is characterized by

\begin{equation}
\kappa
\ll
\delta_b
\ll
1,
\label{eq:extremal_overlap_domain_app}
\end{equation}

for which

\begin{equation}
\chi
\sim
\frac{\delta_b}{\kappa}
\gg
1.
\end{equation}

Equivalently,

\begin{equation}
x
=
\sqrt{3}+\chi
\sim
\frac{\delta_b}{\kappa}
\gg
1.
\end{equation}

In this regime,

\begin{equation}
F_X(x)
=
\frac{
2\sqrt{3}
}{
x
}
+
\Order(x^{-3}),
\end{equation}

while

\begin{equation}
D_x
=
\sqrt{x^2-3}
=
x
\left[
1+\Order(x^{-2})
\right].
\end{equation}

Expanding the matched result
Eq.~\eqref{eq:Phi_match_app} therefore gives

\begin{equation}
\Phi
\sim
\frac{
2\sqrt{3}
}{
\delta_b
}
-
\frac{
4\sqrt{3}
}{
9
}
\ln\delta_b
+
C_\phi
+
\cdots.
\label{eq:extremal_overlap_app}
\end{equation}

Thus the uniform near-extremal solution approaches the characteristic
power-law plus logarithmic structure of the exactly extremal
prograde Kerr strong-deflection problem
\cite{FeleppaBozzaDeGraaf2026}.

\subsection{Comparison with the exact-extremal Kerr expansion}

The overlap result in Eq.~\eqref{eq:extremal_overlap_app} can be
compared directly with the independent exactly extremal
strong-deflection expansion of
Ref.~\cite{FeleppaBozzaDeGraaf2026}. This provides a nontrivial
endpoint check of the leading power-law coefficient, the logarithmic
coefficient, and the finite matching constant.

Reference~\cite{FeleppaBozzaDeGraaf2026} uses units \(2M=1\) and
introduces the near-critical turning-point variable

\begin{equation}
\delta_{\rm F}
=
r_0-r_H,
\qquad
r_H
=
\frac{1}{2},
\label{eq:delta_F_app}
\end{equation}

where the radial coordinate is expressed in units of \(2M\).
Restoring the mass scale gives

\begin{equation}
\delta_{\rm F}
=
\frac{
r_0-M
}{
2M
}.
\label{eq:delta_F_restore_app}
\end{equation}

At exact extremality,

\begin{equation}
a
=
M,
\qquad
\Delta
=
(r-M)^2.
\end{equation}

The exact prograde turning-point relation
Eq.~\eqref{eq:turning_app_exact} then becomes

\begin{align}
b
&=
M+
\frac{
r_0^2
}{
M+(r_0-M)
}
\nonumber\\
&=
M+r_0.
\label{eq:extremal_b_r0_app}
\end{align}

It follows that

\begin{equation}
\delta_b
=
\frac{
b-2M
}{
M
}
=
\frac{
r_0-M
}{
M
}
=
2\delta_{\rm F}.
\label{eq:delta_conversion_app}
\end{equation}

For extremal Kerr,
Ref.~\cite{FeleppaBozzaDeGraaf2026} obtains the total azimuthal shift
\(\Phi_{\rm ext}=\hat{\alpha}_{\rm ext}+\pi\) in the form

\begin{align}
\Phi_{\rm ext}
={}&
\frac{
\sqrt{3}
}{
\delta_{\rm F}
}
-
\frac{
4\sqrt{3}
}{
9
}
\ln\delta_{\rm F}
+
C_{\rm F}
+
\Order(\delta_{\rm F}),
\label{eq:Feleppa_form_app}
\end{align}

where

\begin{equation}
C_{\rm F}
=
\frac{4}{9}
\left[
\sqrt{3}
\ln
\left(
18(2-\sqrt{3})
\right)
+
3(\sqrt{3}-1)
\right].
\label{eq:CF_app}
\end{equation}

Using
\(\delta_{\rm F}=\delta_b/2\),
Eq.~\eqref{eq:Feleppa_form_app} becomes

\begin{align}
\Phi_{\rm ext}
={}&
\frac{
2\sqrt{3}
}{
\delta_b
}
-
\frac{
4\sqrt{3}
}{
9
}
\ln\delta_b
\nonumber\\
&+
C_{\rm F}
+
\frac{
4\sqrt{3}
}{
9
}
\ln2
+
\Order(\delta_b).
\label{eq:Feleppa_converted_app}
\end{align}

The finite contribution satisfies

\begin{align}
C_{\rm F}
+
\frac{
4\sqrt{3}
}{
9
}
\ln2
&=
\frac{
4\sqrt{3}
}{
9
}
\left[
3+
\ln
\left(
36(2-\sqrt{3})
\right)
\right]
-
\frac{4}{3}
\nonumber\\
&=
C_\phi.
\label{eq:Cphi_Feleppa_match_app}
\end{align}

The extremal endpoint of the present uniform expansion therefore
agrees with the independent exactly extremal Kerr result
coefficient by coefficient:

\begin{equation}
\Phi
\sim
\frac{
2\sqrt{3}
}{
\delta_b
}
-
\frac{
4\sqrt{3}
}{
9
}
\ln\delta_b
+
C_\phi
+
\cdots.
\label{eq:extremal_coefficient_match_app}
\end{equation}

The agreement of the power-law coefficient, logarithmic coefficient,
and finite constant provides an independent endpoint check of the
near--outer matching carried out above.

\subsection{Next-to-leading-order image position}

For the \(n\)th image branch,

\begin{equation}
\Lcal_n
=
2\pi n+\psi,
\end{equation}

with \(n\in\mathbb{N}\) and \(\psi=\Order(1)\), as defined in
Sec.~\ref{sec:images}.

We define the physical scaled image position by

\begin{equation}
x_n
\equiv
\sqrt{3}+\chi_n
=
\sqrt{3}
+
\frac{
b_n-b_c
}{
M\kappa
}.
\label{eq:xn_physical_app}
\end{equation}

To determine its next-to-leading-order correction, write

\begin{equation}
x_n
=
x_n^{(0)}
+
\delta x_n,
\qquad
\delta x_n=o(1).
\label{eq:xn_expansion_app}
\end{equation}

At leading order, the image condition gives

\begin{equation}
F_X(x_n^{(0)})
=
\kappa\Lcal_n.
\label{eq:image_leading_equation_app}
\end{equation}

Inverting Eq.~\eqref{eq:F_result_app} yields

\begin{equation}
x_n^{(0)}
=
\sqrt{3}
\coth
\left(
\frac{
\kappa\Lcal_n
}{
2
}
\right).
\label{eq:x0_app}
\end{equation}

Define

\begin{equation}
D_n
\equiv
\sqrt{
(x_n^{(0)})^2-3
}.
\end{equation}

Using Eq.~\eqref{eq:x0_app},

\begin{equation}
D_n
=
\sqrt{3}
\operatorname{csch}
\left(
\frac{
\kappa\Lcal_n
}{
2
}
\right).
\label{eq:Dn_app}
\end{equation}

Writing the NLO phase as

\begin{equation}
\Phi
=
\frac{
F_X(x)
}{
\kappa
}
+
G(x,\kappa)
+
o(1),
\end{equation}

with

\begin{equation}
G(x,\kappa)
=
-
\frac{
4\sqrt{3}
}{
9
}
\ln
\left(
\kappa\sqrt{x^2-3}
\right)
+
C_\phi,
\label{eq:G_image_app}
\end{equation}

we substitute
\(x_n=x_n^{(0)}+\delta x_n\) into the image condition
\(\Phi=\Lcal_n\). Expanding about \(x_n^{(0)}\) and using
Eq.~\eqref{eq:image_leading_equation_app} gives

\begin{equation}
\frac{
\delta x_n
}{
\kappa
}
F_X'(x_n^{(0)})
+
G(x_n^{(0)},\kappa)
=
o(1).
\label{eq:image_NLO_condition_app}
\end{equation}

Since

\begin{equation}
F_X'(x_n^{(0)})
=
-
\frac{
2\sqrt{3}
}{
D_n^2
},
\end{equation}

the scaled image position is

\begin{equation}
x_n
=
x_n^{(0)}
+
\kappa
\frac{
D_n^2
}{
2\sqrt{3}
}
\left[
C_\phi
-
\frac{
4\sqrt{3}
}{
9
}
\ln(\kappa D_n)
\right]
+
o(\kappa).
\label{eq:image_NLO_app}
\end{equation}

The logarithmic dependence of the correction on \(\kappa\) is written
explicitly in Eq.~\eqref{eq:image_NLO_app}; thus no
\(\kappa\)-independent coefficient \(x_n^{(1)}\) is introduced.

Using the exact relation

\begin{equation}
\frac{
b_n-b_c
}{
M
}
=
\kappa
\left(
x_n-\sqrt{3}
\right),
\end{equation}

the corresponding next-to-leading-order image impact parameter is

\begin{align}
\frac{
b_n-b_c
}{
M
}
={}&
\kappa
\left(
x_n^{(0)}-\sqrt{3}
\right)
\nonumber\\
&+
\kappa^2
\frac{
D_n^2
}{
2\sqrt{3}
}
\left[
C_\phi
-
\frac{
4\sqrt{3}
}{
9
}
\ln(\kappa D_n)
\right]
+
o(\kappa^2).
\label{eq:bn_NLO_app}
\end{align}

The first term in Eq.~\eqref{eq:bn_NLO_app} reproduces the leading
uniform image law, while the second term gives the matched NLO
correction.

Equation~\eqref{eq:image_NLO_app} follows from the fixed-\(\chi\)
next-to-leading-order expansion. It is therefore controlled in the
joint near-extremal limit at fixed positive scaled phase
\(\kappa\Lcal_n\), equivalently at fixed
\(x_n^{(0)}>\sqrt{3}\).

An arbitrarily deep limit
\(\kappa\Lcal_n\to\infty\) drives
\(x_n^{(0)}\to\sqrt{3}\) and hence the image crossover coordinate
toward the nested \(\chi_n\to0\) region, where the additional
fixed-subextremal boundary-layer analysis discussed in
Sec.~\ref{sec:nlo} is required.

On the extremal-like side,
\(\kappa\Lcal_n\ll1\), the strong-deflection interpretation also
requires

\begin{equation}
\Lcal_n
\gg
1.
\end{equation}

The controlled algebraic image overlap is therefore

\begin{equation}
1
\ll
\Lcal_n
\ll
\kappa^{-1},
\label{eq:image_NLO_domain_app}
\end{equation}

in agreement with the image-domain analysis of
Sec.~\ref{sec:images}.


\section{Coordinate-Time Matching and Image Delays}
\label{app:time}

This appendix derives the regularized coordinate-time expression and
the image-to-image delay law used in Sec.~\ref{sec:time}. The
calculation parallels the azimuthal near--outer matching of
Appendix~\ref{app:nlo}, but the absolute coordinate time contains an
additional universal divergence at spatial infinity that must first
be removed.

\subsection{Large-radius regularization}

The exact outgoing coordinate-time integrand is given by
Eq.~\eqref{eq:dt_dr_app}. At large radius,

\begin{equation}
\frac{\dd t}{\dd r}
=
1
+
\frac{2M}{r}
+
\Order(r^{-2}).
\label{eq:dt_large_r_app}
\end{equation}

The corresponding two-leg scattering time therefore contains the
universal large-radius divergence

\begin{equation}
2R_{\rm max}
+
4M
\ln
\left(
\frac{
R_{\rm max}
}{
M
}
\right).
\label{eq:time_large_divergence_app}
\end{equation}

We adopt the regularization convention

\begin{align}
T_{\rm reg}
\equiv
\lim_{R_{\rm max}\to\infty}
\Bigg[
&
2
\int_{r_0}^{R_{\rm max}}
\frac{\dd t}{\dd r}
\,\dd r
-
2R_{\rm max}
\nonumber\\
&-
4M
\ln
\left(
\frac{
R_{\rm max}
}{
M
}
\right)
\Bigg].
\label{eq:Treg_app}
\end{align}

Alternative subtraction prescriptions that remove the same
large-radius divergence may shift \(T_{\rm reg}\) by a finite
additive constant. Such a constant cancels from image-to-image
arrival-time differences, provided the same subtraction convention is
applied to every trajectory.

\subsection{Near-horizon temporal scaling}

Substituting the joint near-extremal and near-critical scaling into
Eq.~\eqref{eq:dt_dr_app} gives

\begin{equation}
\frac{1}{M}
\frac{\dd t}{\dd y}
=
\frac{2}{\kappa}
f_0(y;X)
+
\Order(1),
\label{eq:dt_dy_leading_app}
\end{equation}

where \(f_0\) is defined in Eq.~\eqref{eq:f01_app}. Comparison with
Eq.~\eqref{eq:phi_integrand_NLO_app} therefore yields

\begin{equation}
\dd t
\simeq
2M\,\dd\phi,
\label{eq:dt_dphi_app}
\end{equation}

for the leading near-horizon singular contribution. This relation is
consistent with the limiting horizon angular velocity

\begin{equation}
\Omega_H
\longrightarrow
\frac{1}{2M},
\qquad
\kappa\to0.
\end{equation}

To isolate the finite difference between the temporal and azimuthal
accumulations, we expand

\begin{equation}
\frac{\dd}{\dd y}
\left(
\frac{t}{M}-2\phi
\right)
=
h(y;X)
+
\Order(\kappa),
\label{eq:h_time_app}
\end{equation}

where

\begin{equation}
h(y;X)
=
\frac{
3y^2-2Xy+1
}{
(y^2-1)\sqrt{Q(y;X)}
}.
\label{eq:h_def_app}
\end{equation}

For large \(y\),

\begin{equation}
h(y;X)
=
\frac{\sqrt{3}}{y}
+
\Order(y^{-2}).
\label{eq:h_large_y_app}
\end{equation}

We introduce the same intermediate matching scale used in
Appendix~\ref{app:nlo},

\begin{equation}
\frac{r_{\rm m}}{M}
=
1+\kappa Y,
\qquad
1
\ll
Y
\ll
\kappa^{-1}.
\label{eq:time_matching_window_app}
\end{equation}

The two-leg near-region contribution to the combination
\(T_{\rm reg}/M-2\Phi\) has the large-\(Y\) form

\begin{equation}
2
\int_{y_0}^{Y}
h(y;X)\,\dd y
=
2\sqrt{3}
\ln
\left(
\frac{Y}{D}
\right)
+
C_{\rm near}^{(t)}
+
o(1),
\label{eq:time_near_match_app}
\end{equation}

where

\begin{equation}
D
\equiv
\sqrt{X^2-3}.
\end{equation}

The corresponding outer-region contribution has the structure

\begin{equation}
-2\sqrt{3}
\ln(\kappa Y)
+
C_{\rm out}^{(t)}
+
o(1).
\label{eq:time_outer_match_app}
\end{equation}

Adding the near and outer contributions cancels all dependence on the
arbitrary matching scale \(Y\). As in Appendix~\ref{app:nlo}, the
result must then be recentered on the exact physical critical impact
parameter. We therefore use

\begin{equation}
x
=
\sqrt{3}+\chi,
\qquad
D_x
\equiv
\sqrt{x^2-3}.
\label{eq:Dx_time_app}
\end{equation}

Since

\begin{equation}
X
=
x+\Order(\kappa),
\end{equation}

one has, at fixed \(x>\sqrt{3}\),

\begin{equation}
D
=
D_x+\Order(\kappa).
\end{equation}

Replacing \(D\) by \(D_x\) in terms that already contribute at order
unity therefore changes the expansion only beyond the order retained
here.

The matched regularized coordinate time takes the form

\begin{equation}
\frac{
T_{\rm reg}
}{
M
}
=
2\Phi
-
2\sqrt{3}
\ln(\kappa D_x)
+
C_\Delta
+
o(1),
\label{eq:Treg_match_app}
\end{equation}

where \(C_\Delta\) denotes the finite remainder associated with the
chosen large-radius subtraction convention. For a fixed convention,
\(C_\Delta\) is common to the image branches considered below and
therefore cancels from image-to-image delay differences.

Equation~\eqref{eq:Treg_match_app} reproduces
Eq.~\eqref{eq:Treg_phi} of the main text.

\subsection{Higher-order image delays}

For the \(n\)th image branch,

\begin{equation}
\Lcal_n
=
2\pi n+\psi,
\label{eq:Lcal_time_app}
\end{equation}

with \(n\in\mathbb{N}\) and \(\psi=\Order(1)\), as defined in
Sec.~\ref{sec:images}.

At leading order, the image-position inversion derived in
Appendix~\ref{app:nlo} gives

\begin{equation}
x_n^{(0)}
=
\sqrt{3}
\coth
\left(
\frac{
\kappa\Lcal_n
}{
2
}
\right).
\label{eq:xn_time_app}
\end{equation}

It follows that

\begin{equation}
D_n
\equiv
\sqrt{
(x_n^{(0)})^2-3
}
=
\sqrt{3}
\operatorname{csch}
\left(
\frac{
\kappa\Lcal_n
}{
2
}
\right).
\label{eq:Dn_time_app}
\end{equation}

Using the image condition

\begin{equation}
\Phi_n
=
\Lcal_n,
\end{equation}

Eq.~\eqref{eq:Treg_match_app} gives the regularized arrival time of
the \(n\)th image branch,

\begin{equation}
\frac{
T_n
}{
M
}
=
2\Lcal_n
-
2\sqrt{3}
\ln
\left[
\sqrt{3}\,\kappa
\operatorname{csch}
\left(
\frac{
\kappa\Lcal_n
}{
2
}
\right)
\right]
+
C_\Delta
+
o(1).
\label{eq:Tn_app}
\end{equation}

For two image orders \(n\) and \(m\), define the signed
image-to-image coordinate-time delay by

\begin{equation}
\Delta T_{nm}
\equiv
T_n-T_m.
\label{eq:DeltaT_definition_app}
\end{equation}

Subtracting Eq.~\eqref{eq:Tn_app} between the two branches eliminates
the convention-dependent constant \(C_\Delta\) and gives

\begin{align}
\frac{
\Delta T_{nm}
}{
M
}
={}&
2(\Lcal_n-\Lcal_m)
\nonumber\\
&+
2\sqrt{3}
\ln
\left[
\frac{
\sinh(\kappa\Lcal_n/2)
}{
\sinh(\kappa\Lcal_m/2)
}
\right]
+
o(1).
\label{eq:DeltaT_app}
\end{align}

Equation~\eqref{eq:DeltaT_app} is the uniform image-to-image
coordinate-time-delay law quoted in Eq.~\eqref{eq:DeltaT_nm}. The
additive regularization constant has canceled completely, so the delay
differences are independent of the chosen large-radius subtraction
convention.

The same scaled phase variables
\(\kappa\Lcal_n\) and \(\kappa\Lcal_m\) that organize the image
positions therefore control the corresponding arrival-time hierarchy.

\subsection{Extremal-like high-order limit}

On the extremal-like side of the crossover,

\begin{equation}
\kappa\Lcal_j
\ll
1,
\qquad
j\in\{m,n\},
\end{equation}

while strong deflection requires
\(\Lcal_j\gg1\). The controlled overlap is therefore

\begin{equation}
1
\ll
\Lcal_m,\Lcal_n
\ll
\kappa^{-1}.
\label{eq:time_extremal_domain_app}
\end{equation}

Within this regime,

\begin{equation}
\sinh
\left(
\frac{
\kappa\Lcal_j
}{
2
}
\right)
=
\frac{
\kappa\Lcal_j
}{
2
}
\left[
1+
\Order\!\left(
\kappa^2\Lcal_j^2
\right)
\right].
\label{eq:sinh_small_app}
\end{equation}

Equation~\eqref{eq:DeltaT_app} consequently reduces to

\begin{equation}
\frac{
\Delta T_{nm}
}{
M
}
\simeq
2(\Lcal_n-\Lcal_m)
+
2\sqrt{3}
\ln
\left(
\frac{
\Lcal_n
}{
\Lcal_m
}
\right).
\label{eq:DeltaT_extremal_app}
\end{equation}

For neighboring image orders,

\begin{equation}
\Lcal_{n+1}
-
\Lcal_n
=
2\pi,
\end{equation}

and therefore

\begin{equation}
\frac{
\Delta T_{n+1,n}
}{
M
}
\simeq
4\pi
+
2\sqrt{3}
\ln
\left(
\frac{
\Lcal_{n+1}
}{
\Lcal_n
}
\right),
\qquad
1
\ll
\Lcal_n
\ll
\kappa^{-1}.
\label{eq:neighbor_delay_general_app}
\end{equation}

For the reference choice \(\psi=0\),

\begin{equation}
\Lcal_n
=
2\pi n,
\end{equation}

so that

\begin{equation}
\frac{
\Delta T_{n+1,n}
}{
M
}
\simeq
4\pi
+
2\sqrt{3}
\ln
\left(
\frac{
n+1
}{
n
}
\right).
\label{eq:neighbor_delay_app}
\end{equation}

The logarithmic correction vanishes at large image order, but this
limit must remain within the extremal-like overlap. More precisely,

\begin{equation}
\frac{
\Delta T_{n+1,n}
}{
M
}
\longrightarrow
4\pi,
\qquad
n\to\infty,
\quad
\kappa n\to0.
\label{eq:neighbor_delay_limit_app}
\end{equation}

Equivalently, the relevant ordering is

\begin{equation}
1
\ll
n
\ll
\kappa^{-1}.
\label{eq:neighbor_delay_domain_app}
\end{equation}

Thus neighboring high-order image branches approach the asymptotic
coordinate-time spacing

\begin{equation}
\Delta T_{n+1,n}
\longrightarrow
4\pi M
\end{equation}

only in the joint high-order extremal-like limit.

This result has a direct near-horizon interpretation. Since

\begin{equation}
\Omega_H
\longrightarrow
\frac{1}{2M},
\end{equation}

one additional azimuthal winding of \(2\pi\) corresponds
asymptotically to the coordinate-time increment

\begin{equation}
\frac{
2\pi
}{
\Omega_H
}
\longrightarrow
4\pi M.
\end{equation}

The limit should not be identified with taking \(n\to\infty\) at
fixed nonzero \(\kappa\). In that ordering,
\(\kappa n\) eventually becomes large and the sequence leaves the
controlled extremal-like overlap. The coordinate-time hierarchy
therefore exhibits the same noncommuting high-order and extremal
limits as the image-position crossover derived in
Sec.~\ref{sec:images}.


\section{Numerical Implementation and Convergence Tests}
\label{app:numerics}

The numerical calculations were designed to test the asymptotic
predictions against quantities obtained directly from the exact Kerr
geodesic equations. The exact azimuthal phase and coordinate-time
integrals were evaluated independently of the near-extremal
expansions. For the higher-order calculations, asymptotic expressions
were used only to guide numerical bracketing or provide initial
estimates; the final benchmark quantities were determined from the
exact Kerr equations.

\subsection{Exact trajectory construction}

For each value of \(\kappa\), the exact spin was reconstructed from

\begin{equation}
\frac{a}{M}
=
\sqrt{1-\kappa^2},
\end{equation}

and the exact critical quantities were evaluated from
Eqs.~\eqref{eq:rc_app_exact} and \eqref{eq:bc_app_exact}.

For a prescribed physical crossover coordinate \(\chi\), the impact
parameter was set to

\begin{equation}
b
=
b_c
+
M\kappa\chi.
\label{eq:b_numeric_app}
\end{equation}

The physical outer turning point was obtained by solving

\begin{equation}
R(r_0)
=
0
\end{equation}

for the scattering root satisfying

\begin{equation}
r_0
>
r_c
>
r_+.
\end{equation}

The exact accumulated azimuthal phase was then evaluated from

\begin{equation}
\Phi_{\rm exact}
=
2
\int_{r_0}^{\infty}
\frac{
(b-a)+aP/\Delta
}{
\sqrt{R}
}
\,\dd r.
\label{eq:Phi_exact_app_num}
\end{equation}

The integrand has an integrable square-root singularity at
\(r=r_0\). To regularize this endpoint for numerical quadrature, we
introduce locally

\begin{equation}
r
=
r_0+u^2,
\qquad
\dd r
=
2u\,\dd u.
\label{eq:turning_regularization_app}
\end{equation}

Near \(u=0\), the factor of \(u\) in the transformed measure cancels
the inverse square-root behavior of the original radial integrand,
leaving a finite transformed integrand at the turning point.

The exact Kerr expressions were used throughout the construction of
the benchmark trajectories. The asymptotic formulas therefore enter
only in the approximations being tested, not in the evaluation of the
reference phase itself.

\subsection{Leading and NLO phase tests}

The leading data-collapse test used

\begin{equation}
\kappa
\in
\{
0.05,\,
0.02,\,
0.01,\,
0.005,\,
0.002
\},
\label{eq:kappa_grid_app}
\end{equation}

together with 11 values of \(\chi\) spanning

\begin{equation}
0.1
\leq
\chi
\leq
10,
\end{equation}

for a total of 55 exact Kerr trajectories.

At fixed \(\chi\), the scaled exact phase
\(\kappa\Phi_{\rm exact}\) converges toward the analytic crossover
function \(F(\chi)\) defined in Eq.~\eqref{eq:F_chi} as
\(\kappa\to0\). This is the numerical data-collapse test shown in
Fig.~\ref{fig:data_collapse}.

For the next-to-leading-order test, the relative phase errors were
defined by

\begin{equation}
\epsilon_A
=
\frac{
|\Phi_{\rm exact}-\Phi_A|
}{
|\Phi_{\rm exact}|
},
\qquad
A\in\{\mathrm{LO},\mathrm{NLO}\}.
\label{eq:errors_app}
\end{equation}

The corresponding improvement factor was defined as

\begin{equation}
\mathcal I
=
\frac{
\epsilon_{\rm LO}
}{
\epsilon_{\rm NLO}
}.
\label{eq:improvement_app}
\end{equation}

The improvement factor exceeds unity for all 55 trajectories. Its
minimum, median, and maximum values are approximately

\begin{equation}
6.35,
\qquad
95.1,
\qquad
9.81\times10^2,
\end{equation}

respectively. Thus, throughout the resolved domain
\(0.1\leq\chi\leq10\), the matched NLO approximation lies closer to
the exact Kerr phase than the leading uniform approximation for every
tested trajectory.

As an independent test of the finite contribution generated by the
near--outer matching, we define the effective finite remainder

\begin{align}
C_{\rm eff}
={}&
\Phi_{\rm exact}
-
\frac{1}{\kappa}
\ln
\left(
\frac{x+\sqrt{3}}{x-\sqrt{3}}
\right)
\nonumber\\
&+
\frac{
4\sqrt{3}
}{
9
}
\ln
\left(
\kappa D_x
\right),
\label{eq:Ceff_app}
\end{align}

where

\begin{equation}
x
=
\sqrt{3}+\chi,
\qquad
D_x
=
\sqrt{x^2-3}.
\end{equation}

If the matched expansion is correct, then at fixed \(\chi>0\),

\begin{equation}
C_{\rm eff}
\longrightarrow
C_\phi,
\qquad
\kappa\to0.
\end{equation}

For each of the 11 fixed values of \(\chi\), the corresponding
sequence \(C_{\rm eff}(\kappa)\) converges toward

\begin{equation}
C_\phi
\simeq
2.7208672467,
\end{equation}

in agreement with Eq.~\eqref{eq:Cphi_exact}. This test isolates the
finite NLO matching contribution rather than comparing only the
complete phase.

\subsection{Exact higher-order image roots}

Exact higher-order image impact parameters were determined by solving

\begin{equation}
\Phi_{\rm exact}(b)
-
\Lcal_n
=
0,
\label{eq:image_root_app}
\end{equation}

where

\begin{equation}
\Lcal_n
=
2\pi n+\psi.
\end{equation}

For the numerical image validation reported in this work, we adopt
the reference phase choice \(\psi=0\), so that

\begin{equation}
\Phi_{\rm exact}(b_n)
=
2\pi n,
\qquad
n\in\mathbb{N}.
\end{equation}

The asymptotic image formula was used only to obtain an efficient
initial bracket for each root. The final value of \(b_n\) was
determined from Eq.~\eqref{eq:image_root_app} using the exact Kerr
phase integral in Eq.~\eqref{eq:Phi_exact_app_num}.

The image test was organized using the scaled image-order variable

\begin{equation}
s
=
2\pi\kappa n,
\label{eq:s_numeric_app}
\end{equation}

introduced in Sec.~\ref{sec:images}. We used
\(\kappa=0.02,\,0.01,\,0.005\) and selected integer image orders so
that their scaled coordinates lie close to the four target crossover
values

\begin{equation}
s_{\rm target}
=
0.5,\,
1,\,
2,\,
4.
\label{eq:s_values_app}
\end{equation}

For each target value and each \(\kappa\), the physical image order
was chosen as the nearest integer,

\begin{equation}
n
=
\operatorname{round}
\left(
\frac{
s_{\rm target}
}{
2\pi\kappa
}
\right).
\label{eq:n_target_selection_app}
\end{equation}

The resulting actual scaled coordinate
\(s=2\pi\kappa n\) therefore satisfies

\begin{equation}
\left|
s-s_{\rm target}
\right|
\leq
\pi\kappa.
\label{eq:s_target_error_app}
\end{equation}

Hence the discrete sequence approaches the desired target value as
\(\kappa\to0\), while preserving the physically required integer
image order.

The corresponding image orders were
\(n=\{4,8,16,32\}\) for \(\kappa=0.02\),
\(n=\{8,16,32,64\}\) for \(\kappa=0.01\), and
\(n=\{16,32,64,127\}\) for \(\kappa=0.005\).

For each exact root, the actual scaled coordinate

\begin{equation}
s
=
2\pi\kappa n
\end{equation}

was used in the comparison with the analytic master curve. Thus the
four numerical sequences are approximately fixed in \(s\), rather
than exactly fixed at finite \(\kappa\). As \(\kappa\) decreases
along a target-\(s\) sequence, \(n=\Order(\kappa^{-1})\), realizing
the discrete joint high-order and near-extremal scaling underlying
Eq.~\eqref{eq:image_master}.

The final exact phase residuals satisfy

\begin{equation}
\left|
\Phi_{\rm exact}(b_n)-2\pi n
\right|
<
4\times10^{-20},
\label{eq:image_root_residual_app}
\end{equation}

and are therefore negligible compared with the finite-\(\kappa\)
asymptotic errors.

For all 12 exact roots, the NLO image prediction lies closer to the
exact Kerr impact parameter than the leading prediction. In addition,
all four target-\(s\) sequences converge systematically toward the
uniform image law as \(\kappa\to0\).

This numerical construction should be distinguished from the
extremal-like endpoint regime
\(1\ll n\ll\kappa^{-1}\). The target-\(s\) sequences instead probe
the crossover at \(s=\Order(1)\), where
\(n=\Order(\kappa^{-1})\), and therefore test the master curve rather
than only one of its endpoint limits.

\subsection{Coordinate-time calculation}

The exact regularized coordinate-time integral was evaluated from
Eq.~\eqref{eq:Treg_app}. For each trajectory, the integral was
computed at the three finite outer cutoffs

\begin{equation}
\frac{R_{\rm max}}{M}
=
10^3,\,
10^4,\,
10^5.
\label{eq:Rmax_time_app}
\end{equation}

After subtracting the universal large-radius contribution

\begin{equation}
2R_{\rm max}
+
4M
\ln
\left(
\frac{
R_{\rm max}
}{
M
}
\right),
\end{equation}

the remaining cutoff dependence was extrapolated to
\(R_{\rm max}\to\infty\).

Assuming a leading \(R_{\rm max}^{-1}\) correction, two Richardson
extrapolations were formed from the successive cutoff pairs
\((10^3,10^4)\) and \((10^4,10^5)\). The higher-cutoff extrapolation
was used as the final numerical estimate, while the difference between
the two extrapolations provides an independent check of the remaining
cutoff error.

The same regularization and extrapolation prescription was applied to
every trajectory before image-to-image differences were formed.
Consequently, any additive constant associated with the subtraction
convention cancels from the delay difference.

For consistency with the image-root validation, the time-delay tests
also use \(\psi=0\). Nine exact image-to-image delay comparisons were
performed for the same three values
\(\kappa=0.02,\,0.01,\,0.005\). The integer image branches were paired
so that their scaled coordinates lie near the three target crossover
intervals
\(0.5\rightarrow1\), \(1\rightarrow2\), and
\(2\rightarrow4\). The actual values
\(s_m=2\pi\kappa m\) and \(s_n=2\pi\kappa n\) associated with the
integer image orders were used in the numerical construction.

The exact delays were compared with the uniform prediction of
Eq.~\eqref{eq:DeltaT_app}. We define the relative delay error by

\begin{equation}
\epsilon_T
=
\frac{
\left|
\Delta T_{\rm exact}
-
\Delta T_{\rm uniform}
\right|
}{
\left|
\Delta T_{\rm exact}
\right|
}.
\label{eq:delay_error_app}
\end{equation}

For each of the three target crossover intervals,
\(\epsilon_T\) decreases systematically as \(\kappa\to0\).
To estimate the convergence order, we compare successive values of
the relative delay error along each target crossover sequence. Since
the near-extremality parameter is reduced by a factor of two between
successive calculations, the local convergence-order estimate is

\begin{equation}
p_i
=
\frac{
\ln
\left(
\epsilon_{T,i}/\epsilon_{T,i+1}
\right)
}{
\ln 2
}.
\label{eq:delay_order_estimator_app}
\end{equation}

The resulting estimates, combined over the tested target crossover
sequences, give an average convergence order of approximately

\begin{equation}
p
\simeq
1.91.
\label{eq:delay_convergence_order_app}
\end{equation}

Equivalently, over the numerical range tested, the delay residual is
consistent with the approximate power-law behavior

\begin{equation}
\epsilon_T
\propto
\kappa^{1.91}.
\end{equation}

The measured convergence order is therefore consistent with a dominant
correction close to \(\Order(\kappa^2)\) over the numerical range
considered, without assuming an exact quadratic asymptotic law beyond
that range.

As a separate check of the outer-cutoff extrapolation, the two
Richardson estimates agree to within approximately

\begin{equation}
7.2\times10^{-11}.
\label{eq:richardson_disagreement_app}
\end{equation}

This numerical uncertainty is well below the finite-\(\kappa\)
asymptotic residuals relevant to the comparison with the uniform
delay law.


\bibliography{references}

\end{document}